\PassOptionsToPackage{unicode}{hyperref}
\PassOptionsToPackage{hyphens}{url}
\documentclass[
  11pt,
  letterpaper,
]{article}
\usepackage{xcolor}
\usepackage{amsmath,amssymb}
\usepackage{iftex}
\ifPDFTeX
  \usepackage[T1]{fontenc}
  \usepackage[utf8]{inputenc}
  \usepackage{textcomp} 
\else 
  \usepackage{unicode-math} 
  \defaultfontfeatures{Scale=MatchLowercase}
  \defaultfontfeatures[\rmfamily]{Ligatures=TeX,Scale=1}
\fi
\usepackage{lmodern}
\ifPDFTeX\else
\fi
\IfFileExists{upquote.sty}{\usepackage{upquote}}{}
\IfFileExists{microtype.sty}{
  \usepackage[]{microtype}
  \UseMicrotypeSet[protrusion]{basicmath} 
}{}
\makeatletter
\@ifundefined{KOMAClassName}{
  \IfFileExists{parskip.sty}{%
    \usepackage{parskip}
  }{
    \setlength{\parindent}{0pt}
    \setlength{\parskip}{6pt plus 2pt minus 1pt}}
}{
  \KOMAoptions{parskip=half}}
\makeatother
\usepackage{color}
\usepackage{fancyvrb}

\DefineVerbatimEnvironment{Highlighting}{Verbatim}{commandchars=\\\{\}}
\newenvironment{Shaded}{}{}

\newcommand{\DataTypeTok}[1]{\textcolor[rgb]{0.56,0.13,0.00}{#1}}

\newcommand{\FunctionTok}[1]{\textcolor[rgb]{0.02,0.16,0.49}{#1}}

\newcommand{\KeywordTok}[1]{\textcolor[rgb]{0.00,0.44,0.13}{\textbf{#1}}}

\newcommand{\OtherTok}[1]{\textcolor[rgb]{0.00,0.44,0.13}{#1}}

\newcommand{\StringTok}[1]{\textcolor[rgb]{0.25,0.44,0.63}{#1}}

\usepackage{longtable,booktabs,array}
\usepackage{calc} 
\usepackage{etoolbox}
\makeatletter
\patchcmd\longtable{\par}{\if@noskipsec\mbox{}\fi\par}{}{}
\makeatother
\IfFileExists{footnotehyper.sty}{\usepackage{footnotehyper}}{\usepackage{footnote}}
\makesavenoteenv{longtable}
\usepackage{graphicx}
\makeatletter
\newsavebox\pandoc@box
\newcommand*\pandocbounded[1]{
  \sbox\pandoc@box{#1}%
  \Gscale@div\@tempa{\textheight}{\dimexpr\ht\pandoc@box+\dp\pandoc@box\relax}%
  \Gscale@div\@tempb{\linewidth}{\wd\pandoc@box}%
  \ifdim\@tempb\p@<\@tempa\p@\let\@tempa\@tempb\fi
  \ifdim\@tempa\p@<\p@\scalebox{\@tempa}{\usebox\pandoc@box}%
  \else\usebox{\pandoc@box}%
  \fi%
}
\def\fps@figure{htbp}
\makeatother
\usepackage{etoolbox}
\usepackage{booktabs}
\usepackage{longtable}
\usepackage{graphicx}
\usepackage[left=2.5cm,right=2.5cm,top=2.5cm,bottom=2.5cm]{geometry}
\usepackage{microtype}
\AtBeginEnvironment{longtable}{\footnotesize\setlength{\tabcolsep}{4pt}}

\AtBeginEnvironment{verbatim}{\scriptsize}
\AtBeginEnvironment{Shaded}{\scriptsize}

\usepackage{url}
\usepackage[hidelinks]{hyperref}

\newcommand{\dispurl}[2]{\unskip\newline\url{#1}#2\newline\ignorespaces}

\AtBeginDocument{%
}

\AtBeginDocument{\hypersetup{pdftitle={PaperAtlas: an automatically constructed atlas of computational methods and software from 6.4 million open-access articles},pdfauthor={Neeti Shah, Anagha Vippagunta, Yash Bhargava}}}
\usepackage{bookmark}
\IfFileExists{xurl.sty}{\usepackage{xurl}}{} 
\hypersetup{
  hidelinks,
  pdfcreator={LaTeX via pandoc}}

\author{}
\date{}

\begin{document}

\section{PaperAtlas: an automatically constructed atlas of computational methods and software from 6.4 million open-access articles}\label{paperatlas-an-automatically-constructed-atlas-of-computational-methods-and-software-from-6.4-million-open-access-articles}

Neeti Shah\textsuperscript{1,2}, Anagha Vippagunta\textsuperscript{1}, and Yash Bhargava\textsuperscript{1,3,*}

\textsuperscript{1} Bhargava Systems Research Inc., Carmel, IN, USA

\textsuperscript{2} Cornell University, Ithaca, NY, USA

\textsuperscript{3} Johns Hopkins University, Baltimore, MD, USA

* Corresponding author: Yash Bhargava (yash@bhargavaresearch.org)

ORCID: Neeti Shah 0009-0007-6852-8396; Anagha Vippagunta 0009-0006-2971-6696; Yash Bhargava
0009-0002-4615-1448

\subsection{Abstract}\label{abstract}

Computational methods and software are dispersed across a literature that is increasingly difficult to review manually, while curated registries capture only a subset of available resources. We present PaperAtlas, an automatically constructed atlas derived from the PubMed Central open-access corpus. Of 6,446,741 abstracts screened, 1,074,191 were classified as computational, with schema-valid records extracted for 1,074,140. Among these, 267,893 papers describing an algorithm, software package or web server formed 1,438 clusters, with 1,000 retained after restriction to parent-level biomedical categories. In 296 clusters linked to at least five bio.tools entries, the mean concentration of the most frequent EDAM topic was 61.7\%, compared with 26.4\% under permutation. Of 31,180 distinct software and web-server names, 61.0\% lacked a strict match in bio.tools, PyPI, CRAN, Bioconductor or Bioconda. Among 12,207 bio.tools entries with defining papers in the corpus, 56.6\% were recovered end to end, increasing to 83.6\% among the 8,265 entries whose defining publication entered the atlas. PaperAtlas is fully open source and available as a web server at bhargavaresearch.org/paperatlas.

\textbf{Resource:} \dispurl{https://bhargavaresearch.org/paperatlas}{}

\subsection{Introduction}\label{introduction}

Computational methods are integral to most areas of biological research, and they are published at an increasing rate, which makes them harder to collect and organize (1). Methods are scattered across subfields, named inconsistently, and often reported in papers whose title presents only a biological result. Retrieval systems, from lexical whole-MEDLINE ranking (2) to
machine-learning recommenders (3), rank papers by relevance to a query.

A parallel effort extracts software from the literature, motivated by the finding that only 31-43\%
of software mentions carry a formal citation (4). Softcite supplied a machine-learning-ready collection (5), SoftwareKG extended extraction to more than 3 million PubMed Central (PMC) articles (6), and the largest effort spans roughly 20 million
papers (7). A survey of 713,634 PMC full texts found 70\% of resource names mentioned
exactly once (8). On the curation side, bio.tools is hand-curated (9) and FAIRsoft
aggregates 43,987 software instances across eleven sources (10). Such extractions are used
to identify under-recognised packages (11), and related resources annotate the literature at
comparable scale (12). Registry coverage has been estimated within individual subfields, such as synthetic biology (13). In contrast to mention extraction, PaperAtlas is an effort to use large language models (LLMs) to identify papers in which computation is itself a contribution, extract the artifact each contributes and organize these contributions into navigable topics.

We built PaperAtlas to organize the computational contributions of the open-access literature and to ask the following questions (Figure 1). What computational methods and software are introduced in the literature and how are they distributed across topics? Among contributions of the kind a software registry can hold, how often do they match registry records, and how many registered tools does an automated LLM pipeline recover?

\subsection{Materials and methods}\label{materials-and-methods}

\subsubsection{Corpus acquisition}\label{corpus-acquisition}

We parsed 7,283,594 articles from the PMC bulk Journal Article Tag Suite (JATS) XML distribution (14, 15). Each record retained identifiers, bibliographic metadata, sectioned abstract and body text, and figure and table captions. Of these, 6,446,741 had both a title and an abstract and were screened, and the subset excludes methods published only in subscription journals.

\subsubsection{Model and encoder selection}\label{model-and-encoder-selection}

Screening and extraction both used Qwen2.5-Instruct checkpoints (16). The model family was chosen as it is openly licensed, released from 3B to 32B parameters and supports the constrained decoding that the screening labels and the extraction schema require. Screening, which reads one abstract and emits one token, used the 32B checkpoint; extraction, which reads a longer window and writes a structured record for every paper in scope, used the 7B checkpoint. Papers were embedded for clustering with bge-base-en-v1.5 (17), a 768-dimensional sentence encoder and smaller than the other two encoders we compared under identical data splits and an identical classification head, bge-large-en-v1.5 and e5-large-v2 (18). Table S9 reports both comparisons, across the four model scales and across the three encoders.

\subsubsection{Abstract screening}\label{abstract-screening}

Screening decides which papers enter the atlas, and so we designated a paper as in scope when computation is a central scientific result, including algorithm and model development, software and resource papers, and benchmarking studies. Each abstract was assigned to this computational class or to a primarily experimental class which included experimental, clinical and epidemiological work and reviews in which computation is routine statistical analysis; survival, risk-factor and regression analyses of clinical records, a recurring ambiguous case, have an explicit rule in the prompt. Screening used Qwen2.5-32B-Instruct served with vLLM (19) in bfloat16
at temperature 0, with decoding restricted to the tokens spelling the two labels (20, 21). The prompt is reproduced in Table S1.

\subsubsection{Structured extraction}\label{structured-extraction}

Papers classified as computational underwent further extraction with Qwen2.5-7B-Instruct at temperature 0, using
guided JSON decoding against a strict schema (20). Each prompt included the title, the
abstract up to 1,600 characters and the first 2,296 tokens of the body, which was shorter than the
full body for 1,014,058 records (94.40\%). Each record gives a ten-way artifact
type, the artifact name, biological questions, related tools, databases used and a one-sentence
summary (Table S2). The prompt frames every paper as biomedical, including those later excluded by the biomedical restriction.

\subsubsection{Compiling the atlas \& restriction to biomedical topics}\label{compiling-the-atlas-restriction-to-biomedical-topics}

To compile the atlas, we clustered all 267,893 papers of the three artifact types that describe a method or software: new algorithms (215,061), software packages (42,455) and web servers (10,377), here called tool-type papers. These include 54,205 papers without a recovered artifact name, since clustering does not require one. We excluded
the model class, since only 20.8\% of its 290,327 papers name the model they contribute, against
79.8\% for the included types, and 43.7\% address a clinical prediction question, against 20.4\%.
Database and pipeline papers, which release data or assemble existing tools, were not clustered;
their registry matching is reported separately (Table S12).

Each paper's title, summary and biological questions were embedded with bge-base and L2-normalized,
reduced with uniform manifold approximation and projection (UMAP) (22) to five components (15 neighbors, minimum distance 0, cosine, seed
42), and clustered with hierarchical density-based spatial clustering of applications with noise (HDBSCAN) (23, 24) in scikit-learn (25) at minimum cluster size 20, minimum
samples 5 and excess-of-mass selection. Papers that HDBSCAN left unassigned were placed in the nearest cluster when their cosine similarity to its centroid reached 0.80, and were left unassigned otherwise. We set that threshold by hand at the eighth percentile of the similarity between clustered papers and their own centroid, so a paper joins a cluster only if it sits about as close to it as the least typical papers already there (Table S13). Cluster keywords were derived by class-based term frequency-inverse document frequency (TF-IDF) (26) and cluster names generated by Qwen2.5-7B-Instruct from the keywords and
representative titles. Parent topics were obtained by average-linkage agglomeration of cluster
centroids, cut at the prespecified number of parents that maximized centroid silhouette (Table S13).

This corpus included engineering, physics and computer-science articles, because screening did not attempt to constrain the publication's domain. Qwen3-8B (27) classified each cluster as biomedical or non-biomedical from its label, keywords and twelve representative titles, and a parent topic was retained when more than half of its papers lay in clusters classified as biomedical.

\subsubsection{Validation of the categories}\label{validation-of-the-categories}

Three properties of the partition were measured: geometric separation, how far apart the clusters lie in the embedding, measured by cosine silhouette (28) on 12,000 points and by density-based clustering validation (DBCV) (29) on 10,000 papers. Semantic coherence, whether a cluster's keywords belong together, was measured by a 200-trial word-intrusion test (30) against a chance rate of 0.20. Agreement with an independent labeler was measured against a separate language model's labels for 1,500 papers, by adjusted mutual information (AMI) (31), adjusted Rand index (32) and raw agreement. Stability was assessed by repeating the clustering with other UMAP seeds and settings, other HDBSCAN parameters and without papers lacking a recovered artifact name, comparing each repetition with the reference partition by AMI and by the number of retained clusters overlapping at Jaccard 0.5 or above (33); each cluster's median overlap across seeds is its released stability score (Table S13).

External validation used curated annotations: for each atlas paper that is the defining publication of a bio.tools entry, identity-resolved as described below, we took the topics and operations of the EDAM ontology of bioinformatics concepts
(34) that curators assigned to that entry. The primary analysis used entries that are the
only entry of their publication, to avoid dependence between entries of one paper; a sensitivity
analysis used all linked entries and permuted publications as blocks. For a cluster \(c\) with entry set \(E_c\), the concentration of curated annotation is the share of its entries carrying the term \(t\) that is most frequent within it,

\[\mathrm{conc}(c) = \frac{\max_{t} |\{e \in E_c : t \in T(e)\}|}{|E_c|},\]

where \(T(e)\) is the set of terms curators assigned to entry \(e\). Clusters with at least five entries were evaluated. In each of 1,000 permutations, cluster labels were shuffled among entries and each cluster's most frequent term recomputed; with \(b\) permutations reaching the observed value,

\[p = \frac{1 + b}{1001}.\]

Qwen2.5-32B-Instruct was given the title, abstract and
extracted fields of 300 random papers. Its agreement with the extracted artifact type was 84.3\% and
with the artifact name 94.9\%, a consistency check of the small models and not a measure of accuracy (Table S6).

In a human audit, one annotator labeled stratified random samples of 228 items for four tasks,
screening, artifact type and name against the full article, unmatched names and name matches,
without seeing the screening class or the extracted type. Estimates weight each stratum by its
population, with 95\% intervals from 2,000 bootstrap resamples within strata; unclear labels were excluded from the estimate and reported as bounds (Table S16).

\subsubsection{Registry comparison}\label{registry-comparison}

To assess how the extracted software names relate to existing curated resources, we compared them against established software registries and package ecosystems. Extracted names passed two heuristic filters intended to remove likely non-names: one for names composed entirely of
platform, method and container words, and one for multi-token candidates containing neither a
capital letter nor a digit (Table S4). Names were also compared with bio.tools (9), a curated life-science software registry, and separately with four package ecosystems: PyPI
(https://pypi.org), CRAN (https://cran.r-project.org), Bioconductor (35) and Bioconda
(36). A strict match requires equality after folding case, punctuation and trailing
version tokens, or a match of a parenthetical acronym, and all five were retrieved in September 2026 (Table S11).

Strict matching uses registry names alone, whereas an identity-resolved match to a bio.tools entry also uses the entry's publications and links. A match was accepted if the names matched strictly, if the extracted paper was a defining publication for the entry and the names or reported URLs agreed, or if both the URL and name form matched. Name and URL matching rules are defined in Table S12.

Registry matching is reported by artifact type, publication year and parent topic, and papers classified as software packages or web servers form the core software subset, chosen for specificity; the subset is defined by the model's artifact types, so its denominator inherits artifact-classification error. A sensitivity analysis also included all artifact types considered registrable in bio.tools
(Table S12). New algorithms are reported in two groups: those whose paper gives a repository or package link, and those whose paper gives none.

\subsubsection{Recall against bio.tools}\label{recall-against-bio.tools}

The reference set was derived exclusively from bio.tools, where an entry's defining publications were taken as its Primary publications or, when none were available, its untyped publications, resolved to the corpus by PubMed Central identifier (PMCID), PubMed identifier (PMID) or digital object identifier (DOI) using the PMC ID Converter. Each entry was evaluated using a single defining publication selected
a priori: the earliest defining publication represented in the corpus. The principal analysis set, the software subset, comprises entries carrying a tool type other than Database portal and Ontology. An entry counts as recovered when the name extracted from its selected defining publication matches the entry's registry name, so recall over a set \(R\) of entries is

\[\mathrm{recall}(R) = \frac{|\{r \in R : r \mathrm{\ recovered}\}|}{|R|}.\]

End-to-end recall takes \(R\) to be all 12,207 entries of the software subset, so an entry is a loss whenever its paper was screened as experimental, was extracted as a type that is not clustered, was left unassigned by HDBSCAN, fell outside a retained parent topic, yielded no usable name, or yielded a name that did not match. Recall conditional on entry into the atlas restricts \(R\) to entries whose defining publication reached the retained atlas, thereby excluding upstream losses from the denominator and isolating name-matching performance. Recall estimates are reported with Wilson 95\% confidence intervals (CI) (37) and stratified by publication year, tool type, whether the name appears in the title or abstract, and code-link detection. For extraction-level recall, we included all artifact types designated as registrable in the Table S12 crosswalk (new algorithms, software packages, web servers, databases, pipelines and models), without applying clustering or topic restrictions, and this set was specified before any entry was evaluated.

\subsubsection{Retrieval layer and structured store}\label{retrieval-layer-and-structured-store}

The retrieval layer, which was not used in the primary analyses, indexes the 1,074,191 papers classified as computational as 30,913,457 passages embedded with Qwen3-Embedding-0.6B (38), fusing BM25
(39) and dense retrieval by reciprocal rank fusion (40) with cross-encoder
reranking (41). In a known-item retrieval benchmark using the curated descriptions of 300 bio.tools tools with names masked, it ranked the defining paper first for 82.7\% of queries and in the top ten for 93.7\% (Table S15). Alongside the index, a columnar store (42) holds the papers, 41,890,329 parsed reference entries, dataset accessions detected by 22 identifier patterns, and code links to eight hosts normalized to one repository, project, record or package identity (Tables S8 and S15). The deployed resource also assigns newly indexed bioRxiv (43) and arXiv preprints to their nearest cluster above a manually set threshold, holding the rest as emerging; this is not evaluated here (Table S15).

\subsubsection{Language models and AI assistance}\label{language-models-and-ai-assistance}

All models were used as released: Qwen2.5-32B-Instruct for abstract screening and for the extraction consistency check, Qwen2.5-7B-Instruct for structured extraction and cluster naming, Qwen3-8B for classifying clusters as biomedical, bge-base for the clustering embeddings and Qwen3-Embedding-0.6B for the passage index. Every prompt is reproduced in the supplementary material.

During the preparation of this work the authors used OpenAI ChatGPT (GPT-5) to improve the grammar and readability of the manuscript. After using these tools, the authors reviewed and edited the content as needed and take full responsibility for the content of the published article.

\subsection{Results}\label{results}

\subsubsection{Corpus and screening}\label{corpus-and-screening}

Of 7,283,594 parsed articles, 6,446,741 had a title and an abstract, and the screen classified 1,074,191 (16.66\%) as computational, 5,371,134 (83.32\%) as experimental and 1,416 (0.022\%) with no valid label (Table 1, Figure 1b, 1d). The share classified as computational rose
from 13.5\% of papers published in 2012 to 20.0\% in 2025 (Table S10).

\subsubsection{Structured extraction}\label{structured-extraction-1}

Analysis-only studies formed the largest artifact type at 28.19\%, followed by models at 27.03\% and
new algorithms at 20.02\% (Figure S1a, Table S10), and a named artifact was recovered for 316,138 papers (29.43\%). In 1,000 randomly drawn papers extracted again with body text up to the model's 32,768-token context limit, the longer window held the whole body for 985. Artifact type agreed between fixed-window and long-context extraction for 73.7\% of
papers, against 97.3\% between two repeated fixed-window extractions, and names agreed for 74.4\% of
papers named in both. Long-context extraction gave a name for 48.1\% of papers against 30.2\%, and a tool type for 30.6\% against 25.5\%. Which window is more accurate was not measured, so these counts are specific to the production window (Table S14).

\subsubsection{The atlas and its validation}\label{the-atlas-and-its-validation}

Clustering the 267,893 tool-type papers yielded 1,438 clusters under 12 parent topics, with 40,122
papers (15.0\%) unassigned. The parent-level restriction retained five parents, 1,000 clusters and
165,432 papers, listed with their labels, sizes and keywords in Tables S3 and S5 (Figure 2a, 2b).

The three measures capture different properties of the partition: the cosine silhouette was -0.0033 between clusters and 0.118 between parent topics, while DBCV was negative for the reference
partition and every variant, indicating that the papers do not form sharply separated geometric clusters under this representation; the clusters partition them for navigation. Word-intrusion accuracy
was 0.875, consistent with coherent within-cluster keywords, and agreement with the language-model
annotator was modest, at 0.282 AMI, 0.235 adjusted Rand index and 57.5\% raw agreement (Figure 2d).

Repeating UMAP with five additional random seeds produced 1,435-1,486 clusters, with AMI values of 0.78-0.79 relative to the reference partition. Among retained clusters, 72.2\% had a stability score of at least 0.5. Changing the HDBSCAN parameters had a larger effect on the number of clusters, which ranged from 747 to 2,867. We also repeated the clustering after removing papers for which no artifact name was recovered, and 547 retained clusters still overlapped their original counterparts at Jaccard similarity 0.5 or above, showing that excluding these papers changes the resulting partition. Across reassignment thresholds, the proportion of unmatched core names remained nearly unchanged at 60.8-61.4\%, whereas overall recall of bio.tools entries varied more substantially, from 30.6\% to 65.8\%, because this measure includes failures at every stage of the pipeline. Restricted to bio.tools entries whose defining papers entered the atlas, recall rose to 82.8-84.0\% (Table S13).

We also compared alternative embedding models, input representations and dimensionality-reduction strategies using 15,000 tool-type papers, of which 7,334 had curated EDAM topic annotations. When UMAP was used for dimensionality reduction, EDAM topic concentration was similar across bge-base, bge-large and e5-large, ranging from 0.61 to 0.66. In contrast, applying HDBSCAN to principal components or directly to the unreduced embeddings produced only 2-7 clusters, indicating that these representations did not yield a useful partition at this scale. We also compared the text used to represent each paper, and a title-and-abstract representation produced more topic-coherent local neighborhoods than the atlas representation based on title, summary and biological questions: 62.9\% of nearest linked neighbors shared a curated EDAM topic, compared with 58.2\% for the atlas representation (Table S13). The atlas representation had been specified and fixed before these comparisons, and choosing a different representation afterward based on agreement with EDAM topics would effectively optimize the atlas against the same bio.tools annotations later used for validation. We therefore treated these comparisons as sensitivity analyses rather than using them to select the final representation post hoc.

We next asked whether papers grouped into the same atlas clusters also shared independently curated biological annotations. This analysis included 7,261 bio.tools entries, restricted to publications associated with a single bio.tools entry, that mapped to 296 retained clusters containing at least five linked entries each. Within these clusters, the most common EDAM topic was shared by an average of 61.7\% of entries, with a per-cluster median of 60.0\% and an interquartile range of 48.0-77.0\%. Under random permutation, the corresponding concentration was only 26.4\% (p = 0.001, the smallest attainable value). The most common EDAM operation similarly accounted for 42.4\% of entries within clusters, compared with 18.8\% under permutation (Figure 2c). These results were essentially unchanged under stricter linkage criteria or when all 7,584 linked entries were included and publications, rather than individual entries, were permuted. This validation could only be performed for a subset of the atlas: the remaining 704 retained clusters were smaller and more recent, and 69.6\% of their papers concerned biomedical image and signal processing, compared with 13.2\% in the evaluated clusters. The observed annotation coherence should therefore be interpreted as evidence for the evaluated subset rather than for all retained clusters.

Finally, we tested whether the biomedical filtering strategy affected the atlas. The overall number of retained papers was relatively stable across filtering approaches, although the classification of individual clusters was not. Because the primary rule retains entire parent topics, 13.3\% of retained papers belonged to clusters that the separate cluster-level classifier labeled non-biomedical, and the cluster-level classifications agreed with the parent-level rule for 1,152 of the 1,438 clusters. Despite these differences, downstream name-matching results changed little: the proportion of unmatched extracted name strings ranged from 75.1\% to 77.0\% across filtering strategies, and strict bio.tools recall was similar, at 56.2\% when the cluster-level classifications were used and 57.3\% when no biomedical restriction was applied (Table S13). The choice of biomedical filtering rule therefore affected which individual clusters were retained but had relatively little effect on the aggregate name-matching and recall results.

\subsubsection{Artifact names contributed by the literature}\label{artifact-names-contributed-by-the-literature}

The retained atlas contains 121,844 distinct artifact name strings identified across 129,169 naming events. Most names were reported only once, with 95.8\% contributed by a single paper. By artifact type, 92,110 distinct names were associated with new algorithms, 25,241 with software packages and 6,179 with web servers (Table 2).

\subsubsection{Registry matching in the core software subset}\label{registry-matching-in-the-core-software-subset}

Of 31,180 distinct names extracted from software-package or web-server papers, 61.0\% had no strict match in any of the five registries, and 59.6\% remained unmatched after identity resolution. bio.tools alone matched 29.3\% after identity resolution (Table 2, Figure 3a). Among new algorithms with a detected code link, 69.3\% were unmatched, and this fell to 56.8\% when the linked repository or package included the method name. New algorithms with no software evidence detected were unmatched at 83.2\%, and our evidence does not establish their eligibility for a software registry.

The unmatched share increased with publication year in the core subset, from 53.5\% for names introduced before 2011 to 66.7\% for names introduced from 2022 onward, with values for 2011-2021 ranging from 55.6\% to 60.1\%. The higher share among the newest names is consistent with registration lag, although a cross-sectional comparison with snapshots from a single month cannot separate lag from cohort and field effects. By parent topic it was 48.1\% for biological data analysis and 75.6\% for biomedical image and signal processing, where bio.tools matched 12.4\% of core names (Table S12).

Across all 121,844 atlas names, 77.0\% were unmatched by strict matching and 76.4\% after identity resolution. Across the alternative restrictions, name filters and inclusion rules in Table S4, the unmatched share ranged from 75.1\% to 81.0\%. Nearly two thirds of these names came from algorithms with no software evidence detected, so the all-name figure is not an estimate of registry coverage. Unmatched names should not automatically be interpreted as unregistered software: in a previous large-scale extraction study, only 69.66\% of the 10,000 most frequent extracted mentions were confirmed as software by manual curation, despite a cross-validated F1 of 0.92 (7), and other software-mention datasets likewise contain substantial fractions of non-software entities (44) (Table S16).

Of 34,232 bio.tools entries, 16,475 had a defining paper in the screened corpus. Another 11,044 cited papers outside PMC, and 2,738 cited PMC papers outside the open-access subset. Entries whose papers were outside PMC were more often command-line tools (38.4\% against 25.8\% among included entries) or libraries (24.2\% against 15.4\%), so the evaluated reference set is not representative of all bio.tools entries and is enriched for web applications and more established software (Table S11).

Strict matching recovered 56.6\% of the 12,207 entries in the software subset (95\% CI 55.7-57.5), and among the 3,349 entries with a Primary publication recall was 54.3\% (95\% CI 52.6-55.9; Table 3). Identity resolution increased these estimates to 62.1\% and 60.3\% respectively, but because the 674 additional matches it introduced were not manually validated, we use strict-match recall as the primary estimate. The largest losses were papers extracted as an artifact type that is not clustered (16.5\%), principally databases, models and pipelines, and papers left unassigned by HDBSCAN (14.2\%; Figure 3b), so end-to-end recall reflects both extraction performance and the atlas inclusion criteria. Recall differed most strongly according to whether the registry name appeared in the defining paper's title or abstract: strict-match recall was 61.6\% for the 10,883 entries whose names appeared there, compared with 15.4\% for the 1,316 entries whose names did not, and among the latter group 41.7\% were lost specifically at the name-matching stage. Differences by code-link detection, tool type and publication year were smaller (Table S11). Of the 8,265 entries whose paper entered the atlas, 83.6\% were recovered by strict matching (Table 3).

Extracted types and registry tool types agreed where both describe the same kind of object: web
servers were registered as web applications in 79\% of cases, whereas models diverged most (Table
S12).

\subsubsection{Human audit}\label{human-audit}

A single annotator manually audited stratified samples from four stages of the pipeline against the full-text articles and relevant registry records (Table S16). Among 50 papers classified as computational, 74.0\% (95\% CI 62.0-86.0) treated computation as a central result, so the 1,074,191 screened-positive papers should be interpreted as papers classified as computational rather than papers whose primary contribution is computational. After weighting the three sampling strata by their population sizes, the estimated sensitivity of the screening step was 92.9\% (95\% CI 88.9-96.6). Among 50 name-based registry matches, 55.1\% (95\% CI 36.4-72.4) referred to the same artifact, including 15 of 23 bio.tools matches, 9 of 25 PyPI matches, and the single audited CRAN and Bioconda matches, so name-based matching to package registries was the least reliable audited stage. All 15 audited unmatched names were judged to represent genuine named methods or software, although for 12 the annotator could not determine whether the same artifact was registered under a different name, and none of the remaining three had an identified registry counterpart. In a separate full-text audit of 13 papers, the extracted artifact type agreed with the annotator in 8 cases, while the extracted artifact name agreed in every paper for which a name was present, and all three papers classified by the pipeline as software packages were judged by the annotator to describe new algorithms. These audits are limited by the small sample sizes and the use of a single annotator, resulting in wide confidence intervals: screening precision was evaluated on 50 papers, extraction on 13 papers and unmatched names on 15 cases. Identity resolution, extraction-window effects, code-link detection, repository and accession extraction, cluster labeling and the biomedical restriction were not independently audited.

\subsubsection{Structured resource index}\label{structured-resource-index}

The store holds 499,531 dataset-accession occurrences, 285,615 of them distinct, in 184,701 papers
(17.2\% of the corpus), and 152,097 code-link occurrences in 97,331 papers, resolving to 77,382
distinct Git repositories and 6,588 package, project and archive records (Table S15). Of 400 sampled
repository links, 385 resolved; URL resolution verifies accessibility but not correspondence to the reported artifact.

The citation graph, joined by PMID or DOI, resolves 5,767,772 of the 41,890,329 reference entries to papers in the corpus (Table S15).

\subsection{Discussion}\label{discussion}

PaperAtlas is primarily a literature-navigation resource. It organizes computational-methods papers for
task-oriented browsing, identifies candidate software names for registry curation, and links papers to
reported dataset accessions and code repositories. The registry analyses address a separate question: how
often extracted artifact names match entries in five registry snapshots from September 2026.

The results also highlight the limitations of name-based registry matching. Recall fell to 15.4\% when the
registry name did not appear in the defining paper's title or abstract, indicating that naming differences
alone can prevent otherwise related records from being linked. Mention-based resources (6, 7), curated registries such as bio.tools (9) and FAIRsoft (10), and PaperAtlas
therefore capture complementary views of the computational-methods literature.

\subsection{Conclusion}\label{conclusion}

PaperAtlas organizes 165,432 computational-methods papers from the PubMed Central open-access subset into
1,000 biomedical clusters, providing a reproducible structure for navigating the literature. Where curated
bio.tools annotations permit evaluation, these clusters show substantial concentration of shared EDAM
topics. Among 31,180 distinct names extracted from software-package and web-server papers, 61.0\% had no
strict match in the five evaluated registry snapshots. The pipeline recovered 56.6\% of bio.tools software
entries end to end by strict matching, rising to 83.6\% among entries whose defining publication entered the
atlas. A small manual audit found the extracted artifact names accurate in the cases examined, while
registry name matching was the least reliable audited step, so these results measure concordance between
names reported in the literature and names present in the evaluated registry snapshots rather than
definitive registry membership. Together, these results show that PaperAtlas can organize the
computational-methods literature at scale, while also revealing substantial gaps and naming inconsistencies
between published software and existing registries.

\subsection{Supplementary material}\label{supplementary-material}

Supplementary material accompanies this article.

\subsection{Funding}\label{funding}

This research received no specific grant from any funding agency in the public, commercial or
not-for-profit sectors. This work was conducted at Bhargava Systems Research Inc., a registered
nonprofit organization.

\subsection{Author contributions}\label{author-contributions}

N.S. was involved in formal analysis, investigation, data curation, validation, visualization and
writing the original draft. A.V. was involved in formal analysis, investigation and writing the original draft. Y.B. was involved in conceptualization, methodology, software, resources, supervision and writing (review and editing). All authors read and approved the final version of the manuscript.

\subsection{Conflict of interest}\label{conflict-of-interest}

The authors declare no competing interests.

\subsection{Data availability}\label{data-availability}

The atlas and the retrieval service are free to use without registration at
\dispurl{https://bhargavaresearch.org/paperatlas}{,} and the release files can be downloaded from
\dispurl{https://bhargavaresearch.org/paperatlas/download}{.} The pipeline code is at
\dispurl{https://github.com/Bhargava-Systems-Research-Inc/PaperAtlas}{,} and the analysis code, supplementary
tables, validation source data, atlas, embeddings, registry snapshots and per-paper extracted records
are archived at \dispurl{https://doi.org/10.5281/zenodo.22917574}{.} The passage index and parsed corpus are
available on request and can be regenerated from the PMC open-access subset with the released code, and verbatim text is distributed only where the source licence permits (Table S7).

This article describes the September 2026 release, which is frozen in the archive; the web resource
adds new preprints monthly, so its current counts can differ.

\subsection{Tables}\label{tables}

\textbf{Table 1.} Corpus acquisition, screening and extraction

\begin{longtable}[]{@{}lrr@{}}
\toprule\noalign{}
Stage & Count & \% of previous \\
\midrule\noalign{}
\endhead
\bottomrule\noalign{}
\endlastfoot
PMC full-text articles parsed & 7,283,594 & --- \\
Records with title and abstract (screened) & 6,446,741 & 88.51 \\
Screened computational & 1,074,191 & 16.66 \\
No valid screening label & 1,416 & 0.022 \\
Schema-valid extraction & 1,074,140 & 99.995 \\
Named artifact recovered & 316,138 & 29.43 \\
Tool-type papers clustered & 267,893 & --- \\
Assigned to a cluster & 227,771 & 85.0 \\
Retained after biomedical restriction & 165,432 & 72.6 \\
\end{longtable}

\textbf{Table 2.} Registry matches for 121,844 distinct artifact names, by artifact type. Values are row percentages. S, strict name matching; I, identity resolution using publication and URL evidence; Packages, PyPI, CRAN, Bioconductor or Bioconda; None, no registry match. Code links are repository or package URLs reported in the paper

\begin{longtable}[]{@{}
  >{\raggedright\arraybackslash}p{(\linewidth - 12\tabcolsep) * \real{0.1111}}
  >{\raggedleft\arraybackslash}p{(\linewidth - 12\tabcolsep) * \real{0.1481}}
  >{\raggedleft\arraybackslash}p{(\linewidth - 12\tabcolsep) * \real{0.1481}}
  >{\raggedleft\arraybackslash}p{(\linewidth - 12\tabcolsep) * \real{0.1481}}
  >{\raggedleft\arraybackslash}p{(\linewidth - 12\tabcolsep) * \real{0.1481}}
  >{\raggedleft\arraybackslash}p{(\linewidth - 12\tabcolsep) * \real{0.1481}}
  >{\raggedleft\arraybackslash}p{(\linewidth - 12\tabcolsep) * \real{0.1481}}@{}}
\toprule\noalign{}
\begin{minipage}[b]{\linewidth}\raggedright
Name strings from
\end{minipage} & \begin{minipage}[b]{\linewidth}\raggedleft
Names
\end{minipage} & \begin{minipage}[b]{\linewidth}\raggedleft
bio.tools S
\end{minipage} & \begin{minipage}[b]{\linewidth}\raggedleft
bio.tools I
\end{minipage} & \begin{minipage}[b]{\linewidth}\raggedleft
Packages
\end{minipage} & \begin{minipage}[b]{\linewidth}\raggedleft
None S
\end{minipage} & \begin{minipage}[b]{\linewidth}\raggedleft
None I
\end{minipage} \\
\midrule\noalign{}
\endhead
\bottomrule\noalign{}
\endlastfoot
Software packages and web servers (core subset) & 31,180 & 27.7 & 29.3 & 23.5 & 61.0 & 59.6 \\
Software packages & 25,241 & 25.9 & 27.2 & 25.6 & 61.7 & 60.6 \\
Web servers & 6,179 & 37.0 & 39.7 & 16.2 & 56.4 & 53.9 \\
New algorithms, code link detected & 14,491 & 16.5 & 17.7 & 23.7 & 69.3 & 68.3 \\
New algorithms, link named as the method & 6,199 & 26.2 & 27.8 & 30.9 & 56.8 & 55.5 \\
New algorithms, no software evidence detected & 78,508 & 6.7 & 6.8 & 14.5 & 83.2 & 83.1 \\
All tool types & 121,844 & 12.4 & 13.0 & 17.0 & 77.0 & 76.4 \\
\end{longtable}

\textbf{Table 3.} Recall for the 12,207 bio.tools entries in the software subset, defined as entries with a qualifying tool type and a defining paper in the screened corpus. Compatible types are new algorithms, software packages, web servers, databases, pipelines and models. Rows use different denominators: the first row includes the full software subset, whereas entries whose paper entered the atlas are conditional on successful completion of the preceding pipeline stages. Intervals are Wilson 95\% confidence intervals. Stage-specific losses are shown in Figure 3b, with additional analyses in Table S11

\begin{longtable}[]{@{}
  >{\raggedright\arraybackslash}p{(\linewidth - 6\tabcolsep) * \real{0.2000}}
  >{\raggedleft\arraybackslash}p{(\linewidth - 6\tabcolsep) * \real{0.2667}}
  >{\raggedleft\arraybackslash}p{(\linewidth - 6\tabcolsep) * \real{0.2667}}
  >{\raggedleft\arraybackslash}p{(\linewidth - 6\tabcolsep) * \real{0.2667}}@{}}
\toprule\noalign{}
\begin{minipage}[b]{\linewidth}\raggedright
Measure
\end{minipage} & \begin{minipage}[b]{\linewidth}\raggedleft
n
\end{minipage} & \begin{minipage}[b]{\linewidth}\raggedleft
Strict
\end{minipage} & \begin{minipage}[b]{\linewidth}\raggedleft
Identity-resolved
\end{minipage} \\
\midrule\noalign{}
\endhead
\bottomrule\noalign{}
\endlastfoot
Entries recovered & 12,207 & 56.6 (55.7-57.5) & 62.1 (61.3-63.0) \\
Entries with a Primary publication & 3,349 & 54.3 (52.6-55.9) & 60.3 (58.7-62.0) \\
Entries from single-entry publications & 11,498 & 58.7 & 63.2 \\
Entries whose paper entered the atlas & 8,265 & 83.6 & 91.8 \\
Compatible types admitted, no clustering & 12,207 & 81.3 & 88.9 \\
Name in the paper's title or abstract & 10,883 & 61.6 & 65.7 \\
Name not in the title or abstract & 1,316 & 15.4 & 33.1 \\
\end{longtable}

\subsection{References}\label{references}

\begin{enumerate}
\def\labelenumi{\arabic{enumi}.}
\item
  Bornmann L, Haunschild R, Mutz R. Growth rates of modern science: a latent piecewise growth curve approach to model publication numbers from established and new literature databases. Humanit Soc Sci Commun 2021;8:224. \url{https://doi.org/10.1057/s41599-021-00903-w}
\item
  Fontaine J-F, Barbosa-Silva A, Schaefer M et al.~MedlineRanker: flexible ranking of biomedical literature. Nucleic Acids Res 2009;37:W141--6. \url{https://doi.org/10.1093/nar/gkp353}
\item
  Allot A, Lee K, Chen Q et al.~LitSuggest: a web-based system for literature recommendation and curation using machine learning. Nucleic Acids Res 2021;49:W352--8. \url{https://doi.org/10.1093/nar/gkab326}
\item
  Howison J, Bullard J. Software in the scientific literature: problems with seeing, finding, and using software mentioned in the biology literature. J Assoc Inf Sci Technol 2016;67:2137--55. \url{https://doi.org/10.1002/asi.23538}
\item
  Du C, Cohoon J, Lopez P et al.~Softcite dataset: a dataset of software mentions in biomedical and economic research publications. J Assoc Inf Sci Technol 2021;72:870--84. \url{https://doi.org/10.1002/asi.24454}
\item
  Schindler D, Bensmann F, Dietze S et al.~The role of software in science: a knowledge graph-based analysis of software mentions in PubMed Central. PeerJ Comput Sci 2022;8:e835. \url{https://doi.org/10.7717/peerj-cs.835}
\item
  Istrate A-M, Li D, Taraborelli D et al.~A large dataset of software mentions in the biomedical literature. arXiv 2022. \url{https://doi.org/10.48550/arXiv.2209.00693}
\item
  Duck G, Nenadic G, Filannino M et al.~A survey of bioinformatics database and software usage through mining the literature. PLoS One 2016;11:e0157989. \url{https://doi.org/10.1371/journal.pone.0157989}
\item
  Ison J, Ienasescu H, Chmura P et al.~The bio.tools registry of software tools and data resources for the life sciences. Genome Biol 2019;20:164. \url{https://doi.org/10.1186/s13059-019-1772-6}
\item
  Martín del Pico E, Gelpí JL, Capella-Gutiérrez S. FAIRsoft---a practical implementation of FAIR principles for research software. Bioinformatics 2024;40:btae464. \url{https://doi.org/10.1093/bioinformatics/btae464}
\item
  Brown EM, Druskat S, Hébert-Dufresne L et al.~Biomedical open source software: crucial packages and hidden heroes. PLoS Comput Biol 2026;22:e1014260. \url{https://doi.org/10.1371/journal.pcbi.1014260}
\item
  Wei C-H, Allot A, Lai P-T et al.~PubTator 3.0: an AI-powered literature resource for unlocking biomedical knowledge. Nucleic Acids Res 2024;52:W540--6. \url{https://doi.org/10.1093/nar/gkae235}
\item
  Cai P, Liu S, Zhang D et al.~SynBioTools: a one-stop facility for searching and selecting synthetic biology tools. BMC Bioinformatics 2023;24:152. \url{https://doi.org/10.1186/s12859-023-05281-5}
\item
  National Information Standards Organization. JATS: Journal Article Tag Suite, version 1.4 (ANSI/NISO Z39.96-2024). Baltimore, MD: NISO, 2024. \url{https://doi.org/10.3789/ansi.niso.z39.96-2024}
\item
  Sayers EW, Bolton EE, Fine AM et al.~Database resources of the National Center for Biotechnology Information in 2026. Nucleic Acids Res 2026;54:D20--7. \url{https://doi.org/10.1093/nar/gkaf1060}
\item
  Qwen Team. Qwen2.5 technical report. arXiv 2024. \url{https://doi.org/10.48550/arXiv.2412.15115}
\item
  Xiao S, Liu Z, Zhang P et al.~C-Pack: packed resources for general Chinese embeddings. In: Proceedings of the 47th International ACM SIGIR Conference on Research and Development in Information Retrieval. New York, NY: ACM, 2024, 641--9. \url{https://doi.org/10.1145/3626772.3657878}
\item
  Wang L, Yang N, Huang X et al.~Text embeddings by weakly-supervised contrastive pre-training. arXiv:2212.03533, 2022. \url{https://doi.org/10.48550/arXiv.2212.03533}
\item
  Kwon W, Li Z, Zhuang S et al.~Efficient memory management for large language model serving with PagedAttention. In: Proceedings of the 29th Symposium on Operating Systems Principles. New York, NY: ACM, 2023, 611--26. \url{https://doi.org/10.1145/3600006.3613165}
\item
  Willard BT, Louf R. Efficient guided generation for large language models. arXiv 2023. \url{https://doi.org/10.48550/arXiv.2307.09702}
\item
  Geng S, Josifoski M, Peyrard M et al.~Grammar-constrained decoding for structured NLP tasks without finetuning. In: Proceedings of the 2023 Conference on Empirical Methods in Natural Language Processing. Stroudsburg, PA: ACL, 2023, 10932--52. \url{https://doi.org/10.18653/v1/2023.emnlp-main.674}
\item
  McInnes L, Healy J, Melville J. UMAP: uniform manifold approximation and projection for dimension reduction. arXiv 2018. \url{https://doi.org/10.48550/arXiv.1802.03426}
\item
  Campello RJGB, Moulavi D, Sander J. Density-based clustering based on hierarchical density estimates. In: Advances in Knowledge Discovery and Data Mining. Lecture Notes in Computer Science, vol.~7819. Berlin: Springer, 2013, 160--72. \url{https://doi.org/10.1007/978-3-642-37456-2_14}
\item
  McInnes L, Healy J, Astels S. hdbscan: hierarchical density based clustering. J Open Source Softw 2017;2:205. \url{https://doi.org/10.21105/joss.00205}
\item
  Pedregosa F, Varoquaux G, Gramfort A et al.~Scikit-learn: machine learning in Python. J Mach Learn Res 2011;12:2825--30.
\item
  Grootendorst M. BERTopic: neural topic modeling with a class-based TF-IDF procedure. arXiv 2022. \url{https://doi.org/10.48550/arXiv.2203.05794}
\item
  Yang A, Li A, Yang B et al.~Qwen3 technical report. arXiv:2505.09388, 2025. \url{https://doi.org/10.48550/arXiv.2505.09388}
\item
  Rousseeuw PJ. Silhouettes: a graphical aid to the interpretation and validation of cluster analysis. J Comput Appl Math 1987;20:53--65. \url{https://doi.org/10.1016/0377-0427(87)90125-7}
\item
  Moulavi D, Jaskowiak PA, Campello RJGB et al.~Density-based clustering validation. In: Proceedings of the 2014 SIAM International Conference on Data Mining. Philadelphia: SIAM, 2014, 839--47. \url{https://doi.org/10.1137/1.9781611973440.96}
\item
  Chang J, Gerrish S, Wang C et al.~Reading tea leaves: how humans interpret topic models. In: Advances in Neural Information Processing Systems 22. Red Hook, NY: Curran Associates, 2009, 288--96.
\item
  Vinh NX, Epps J, Bailey J. Information theoretic measures for clusterings comparison: variants, properties, normalization and correction for chance. J Mach Learn Res 2010;11:2837--54.
\item
  Hubert L, Arabie P. Comparing partitions. J Classif 1985;2:193--218. \url{https://doi.org/10.1007/BF01908075}
\item
  Hennig C. Cluster-wise assessment of cluster stability. Comput Stat Data Anal 2007;52:258--71. \url{https://doi.org/10.1016/j.csda.2006.11.025}
\item
  Ison J, Kalaš M, Jonassen I et al.~EDAM: an ontology of bioinformatics operations, types of data and identifiers, topics and formats. Bioinformatics 2013;29:1325--32. \url{https://doi.org/10.1093/bioinformatics/btt113}
\item
  Huber W, Carey VJ, Gentleman R et al.~Orchestrating high-throughput genomic analysis with Bioconductor. Nat Methods 2015;12:115--21. \url{https://doi.org/10.1038/nmeth.3252}
\item
  Gruning B, Dale R, Sjodin A et al.~Bioconda: sustainable and comprehensive software distribution for the life sciences. Nat Methods 2018;15:475-6. \url{https://doi.org/10.1038/s41592-018-0046-7}
\item
  Wilson EB. Probable inference, the law of succession, and statistical inference. J Am Stat Assoc 1927;22:209--12. \url{https://doi.org/10.1080/01621459.1927.10502953}
\item
  Zhang Y, Li M, Long D et al.~Qwen3 Embedding: advancing text embedding and reranking through foundation models. arXiv:2506.05176, 2025. \url{https://doi.org/10.48550/arXiv.2506.05176}
\item
  Robertson S, Zaragoza H. The probabilistic relevance framework: BM25 and beyond. Found Trends Inf Retr 2009;3:333-89. \url{https://doi.org/10.1561/1500000019}
\item
  Cormack GV, Clarke CLA, Buettcher S. Reciprocal rank fusion outperforms Condorcet and individual rank learning methods. In: Proceedings of the 32nd International ACM SIGIR Conference on Research and Development in Information Retrieval. New York, NY: ACM, 2009, 758-9. \url{https://doi.org/10.1145/1571941.1572114}
\item
  Chen J, Xiao S, Zhang P et al.~M3-Embedding: multi-linguality, multi-functionality, multi-granularity text embeddings through self-knowledge distillation. In: Findings of the Association for Computational Linguistics: ACL 2024. Association for Computational Linguistics, 2024, 2318-35. \url{https://doi.org/10.18653/v1/2024.findings-acl.137}
\item
  Raasveldt M, Muhleisen H. DuckDB: an embeddable analytical database. In: Proceedings of the 2019 International Conference on Management of Data. New York, NY: ACM, 2019, 1981-4. \url{https://doi.org/10.1145/3299869.3320212}
\item
  Sever R, Roeder T, Hindle S et al.~bioRxiv: the preprint server for biology. bioRxiv 2019. \url{https://doi.org/10.1101/833400}
\item
  Druskat S, Chue Hong NP, Buzzard S et al.~Don't mention it: an approach to assess challenges to using software mentions for citation and discoverability research. arXiv 2024. \url{https://doi.org/10.48550/arXiv.2402.14602}
\end{enumerate}

\subsection{Figures}\label{figures}

\pandocbounded{\includegraphics[keepaspectratio,alt={Figure 1}]{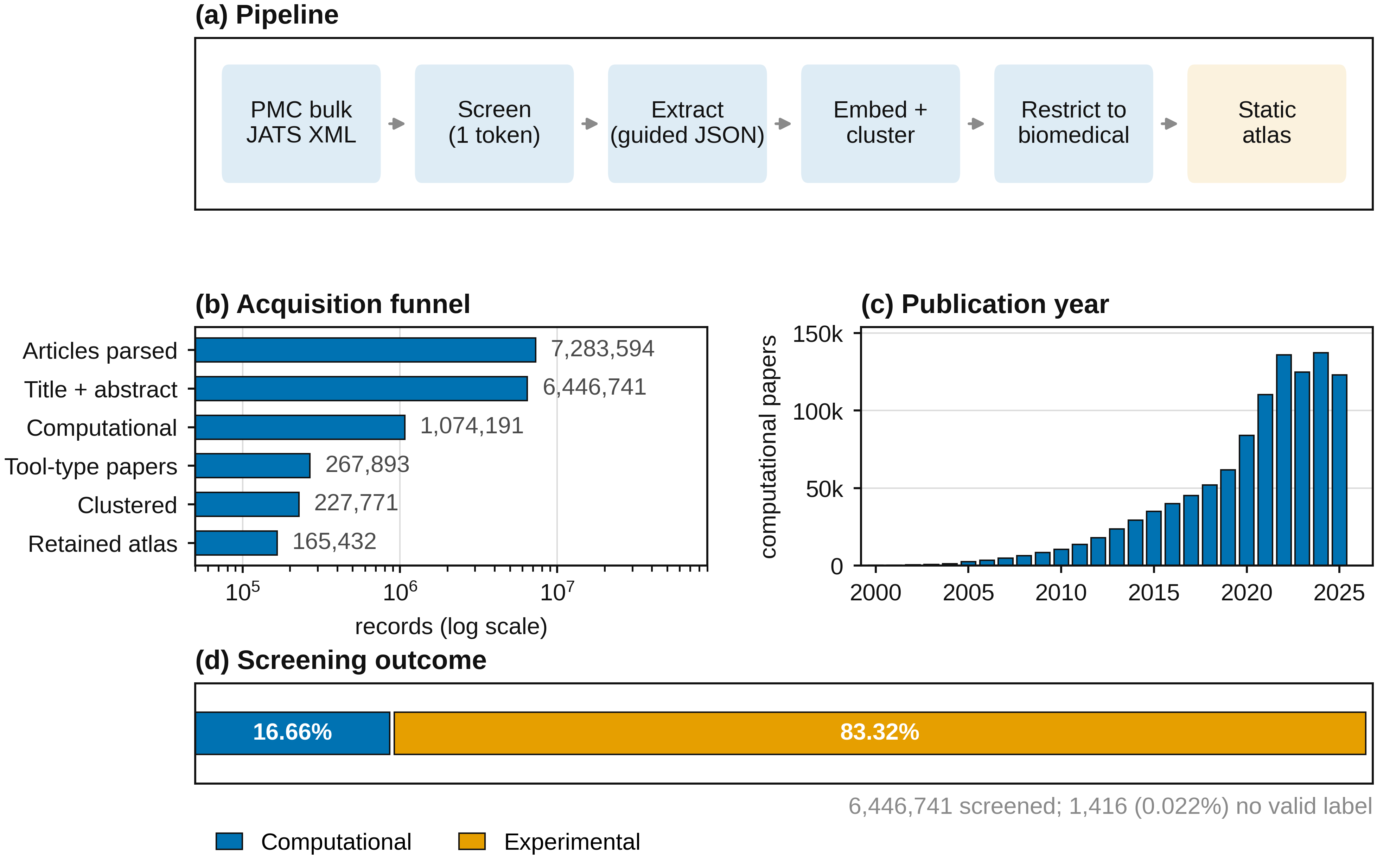}}

\textbf{Figure 1.} The PaperAtlas pipeline and corpus. (a) End-to-end workflow. (b) Acquisition and
screening funnel, 7,283,594 articles parsed to 165,432 retained under the parent-level biomedical rule. (c) Publication year of the
computational corpus. (d) Screening outcome.

\pandocbounded{\includegraphics[keepaspectratio,alt={Figure 2}]{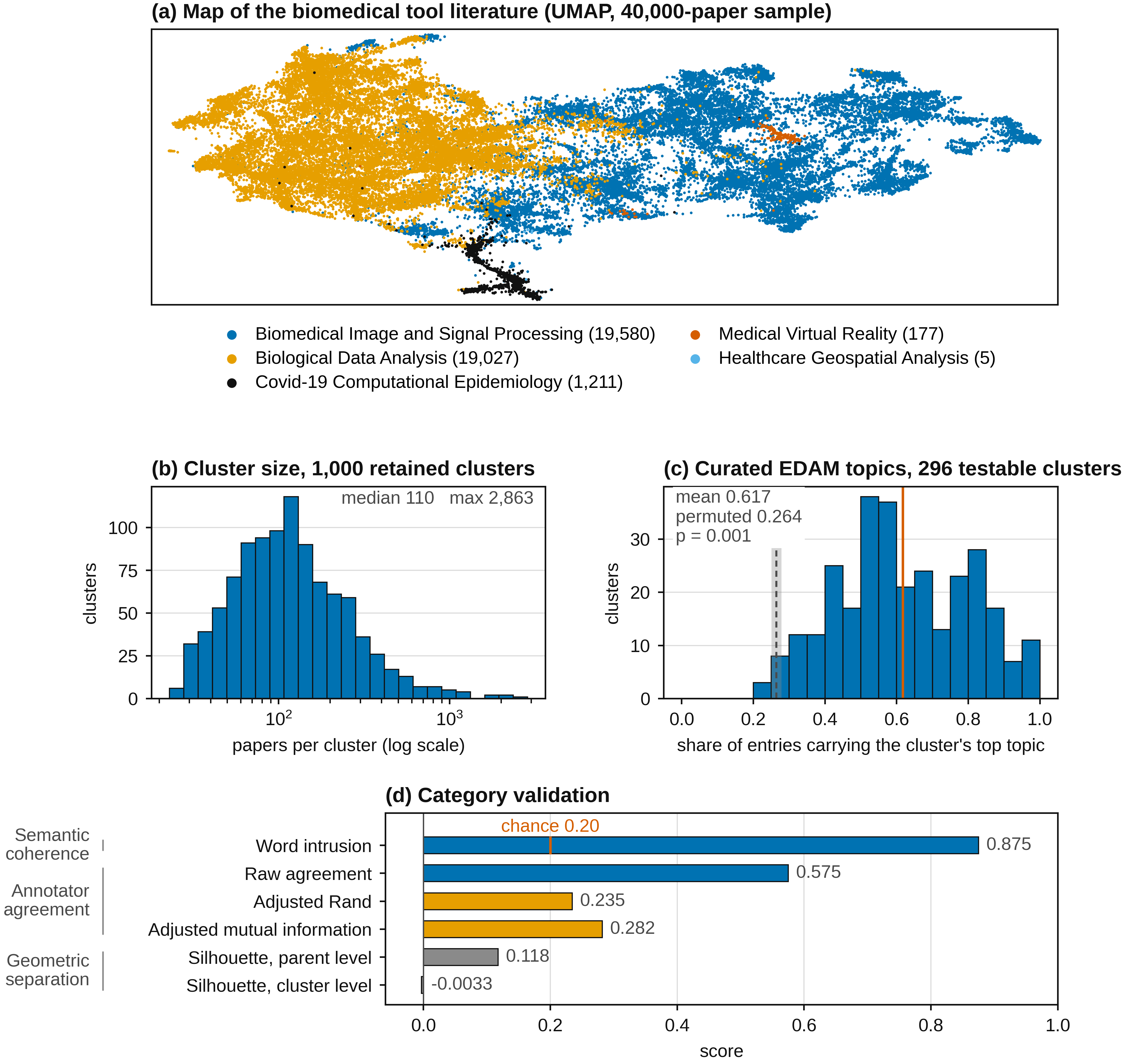}}

\textbf{Figure 2.} The atlas and its validation. (a) UMAP of the retained tool-type paper embeddings,
colored by parent topic, a navigational grouping. (b) Cluster sizes across the 1,000 retained
clusters. (c) For the 296 clusters with at least five linked bio.tools entries, the share of entries
carrying the cluster's most frequent curated EDAM topic. The solid line is the mean, the dashed line
and grey band the mean and range of 1,000 permutations of cluster labels. (d) Validation measures, grouped by the property each tests. Word intrusion is an accuracy with a 0.20 chance floor,
marked. The adjusted Rand index and adjusted mutual information are chance-corrected, so their chance
value is zero. The silhouette ranges from -1 to 1, and zero denotes overlapping clusters. None of these
measures uses human labels.

\pandocbounded{\includegraphics[keepaspectratio,alt={Figure 3}]{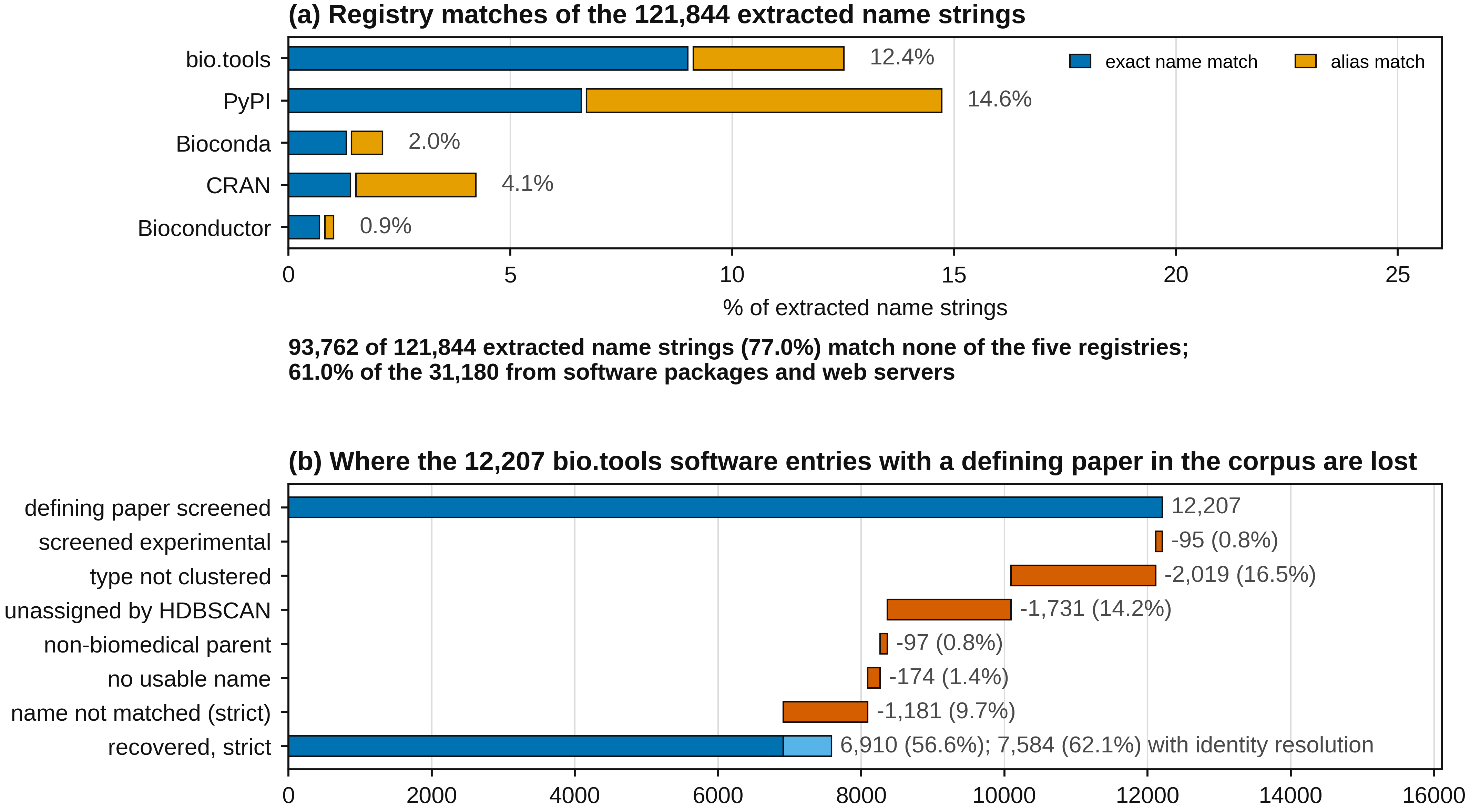}}

\textbf{Figure 3.} Registry matching and bio.tools recall. (a) Exact and alias match rates of the 121,844
extracted name strings against each of the five evaluated registries, with the share matching none for
all name strings and for the core software subset. (b) Where the 12,207 bio.tools software entries with a defining paper in the corpus are lost, stage by stage, ending
with the entries recovered by strict matching and, lighter, those added by identity resolution
(Table S11).

\section{Supplementary Material}\label{supplementary-material-1}

\textbf{PaperAtlas: an automatically constructed atlas of computational methods and software from 6.4
million open-access articles}

Neeti Shah, Anagha Vippagunta and Yash Bhargava

\subsection{Supplementary tables}\label{supplementary-tables}

\textbf{Table S1.} Abstract-screening prompt, reproduced verbatim. Decoding is restricted to the token
identifiers spelling \texttt{C} and \texttt{E}, so no other output is reachable

\begin{verbatim}
Classify the paper's main contribution. C means computational work is a central scientific
result: algorithm/software, simulation, mathematical or mechanistic model, machine learning,
bioinformatics, genome/transcriptome/proteome data analysis, sequence/phylogenetic/structure
analysis, in-silico screening, or large-scale quantitative modeling. E means the main work is
wet-lab, assay, animal, clinical, review, survey, registry/claims/cohort, imaging, or epidemiology
where computation is routine statistics or support analysis. For survival/risk-factor/regression
analyses of clinical records, choose E unless a new computational method/model is the main result.
Answer exactly one letter: C or E.
Title: {title}
Abstract: {abstract}
Answer:
\end{verbatim}

\textbf{Table S2.} The extraction prompt and the JSON Schema enforced during sampling, both reproduced
verbatim. A structurally invalid record cannot be produced

Extraction prompt, reproduced verbatim. Each prompt is followed by the title, the abstract up to
1,600 characters and the opening 2,296 tokens of the body

\begin{verbatim}
You analyze a biomedical paper and return ONE JSON object classifying its computational
contribution for a searchable methods database. Use only the paper's content. Fill EVERY field.

artifact_type — the SINGLE primary thing the authors built/contributed:
  new_algorithm    = a novel method/algorithm is the central result (even with no released code)
  software_package = a reusable library/toolkit/standalone program (e.g. an R/Python package)
  database         = a curated data resource/repository the authors release
  benchmark        = a dataset + evaluation built to compare methods
  pipeline         = an assembled multi-step workflow made mostly from existing tools
  web_server       = an interactive online tool / web application
  model            = a trained ML / statistical / mechanistic model is the main deliverable
  review           = review, survey, perspective, or commentary (no new method/data)
  analysis_only    = applies existing tools to data to answer a biological question; no new tool
                     or resource released
  other            = none of the above
artifact_name — name of that artifact exactly as written, or "" if unnamed.
biological_questions — ALL biological problems addressed (array, may be empty); choose from:
  biomarker_discovery, disease_subtyping, target_discovery, drug_repurposing, diagnosis,
  prognosis, mechanism_inference, patient_stratification, variant_interpretation,
  drug_response_prediction, other, none
  Choose these only when they are the stated biological/clinical purpose of the work.
  Do not label generic protein/domain/gene classification as diagnosis. Do not label generic
  binding-site, structure, or pharmacokinetic work as drug_response_prediction unless the paper
  explicitly predicts response to a drug/treatment. Use ["none"] for generic computational
  methods/resources with no clear listed biological question.
tool_relationships — for OTHER named software/tools the paper relates to, array of
  {"tool": <name>, "relationship": <one of: uses, extends, wraps, compares_against, replaces,
  integrates_with, visualizes, depends_on>}. Empty array if none.
databases_used — external biological databases/resources relied on (e.g. GEO, TCGA, ClinVar,
  GWAS Catalog, DrugBank, ChEMBL, KEGG, Reactome, STRING, UniProt, PDB, Ensembl). Do not copy
  examples; include a database only if its name appears in the paper. Empty if none.
Do not force clinical labels. Use ["none"] when the paper has no clear listed biological question.
summary — one sentence (<=30 words) stating the main computational contribution.

Return ONLY the JSON object.
\end{verbatim}

\textbf{Schema enforced during sampling,} reproduced verbatim. A structurally invalid record is
unreachable, and is never rejected after the fact

\begin{Shaded}
\begin{Highlighting}[]
\FunctionTok{\{}
  \DataTypeTok{"type"}\FunctionTok{:} \StringTok{"object"}\FunctionTok{,} \DataTypeTok{"additionalProperties"}\FunctionTok{:} \KeywordTok{false}\FunctionTok{,}
  \DataTypeTok{"properties"}\FunctionTok{:} \FunctionTok{\{}
    \DataTypeTok{"artifact\_type"}\FunctionTok{:} \FunctionTok{\{}\DataTypeTok{"type"}\FunctionTok{:} \StringTok{"string"}\FunctionTok{,} \DataTypeTok{"enum"}\FunctionTok{:} \OtherTok{[}\StringTok{"new\_algorithm"}\OtherTok{,} \StringTok{"software\_package"}\OtherTok{,} \StringTok{"database"}\OtherTok{,}
        \StringTok{"benchmark"}\OtherTok{,} \StringTok{"pipeline"}\OtherTok{,} \StringTok{"web\_server"}\OtherTok{,} \StringTok{"model"}\OtherTok{,} \StringTok{"review"}\OtherTok{,} \StringTok{"analysis\_only"}\OtherTok{,} \StringTok{"other"}\OtherTok{]}\FunctionTok{\},}
    \DataTypeTok{"artifact\_name"}\FunctionTok{:} \FunctionTok{\{}\DataTypeTok{"type"}\FunctionTok{:} \StringTok{"string"}\FunctionTok{\},}
    \DataTypeTok{"biological\_questions"}\FunctionTok{:} \FunctionTok{\{}\DataTypeTok{"type"}\FunctionTok{:} \StringTok{"array"}\FunctionTok{,} \DataTypeTok{"items"}\FunctionTok{:} \FunctionTok{\{}\DataTypeTok{"type"}\FunctionTok{:} \StringTok{"string"}\FunctionTok{,} \DataTypeTok{"enum"}\FunctionTok{:}
        \OtherTok{[}\StringTok{"biomarker\_discovery"}\OtherTok{,} \StringTok{"disease\_subtyping"}\OtherTok{,} \StringTok{"target\_discovery"}\OtherTok{,} \StringTok{"drug\_repurposing"}\OtherTok{,}
         \StringTok{"diagnosis"}\OtherTok{,} \StringTok{"prognosis"}\OtherTok{,} \StringTok{"mechanism\_inference"}\OtherTok{,} \StringTok{"patient\_stratification"}\OtherTok{,}
         \StringTok{"variant\_interpretation"}\OtherTok{,} \StringTok{"drug\_response\_prediction"}\OtherTok{,} \StringTok{"other"}\OtherTok{,} \StringTok{"none"}\OtherTok{]}\FunctionTok{\}\},}
    \DataTypeTok{"tool\_relationships"}\FunctionTok{:} \FunctionTok{\{}\DataTypeTok{"type"}\FunctionTok{:} \StringTok{"array"}\FunctionTok{,} \DataTypeTok{"items"}\FunctionTok{:} \FunctionTok{\{}\DataTypeTok{"type"}\FunctionTok{:} \StringTok{"object"}\FunctionTok{,}
        \DataTypeTok{"additionalProperties"}\FunctionTok{:} \KeywordTok{false}\FunctionTok{,}
        \DataTypeTok{"properties"}\FunctionTok{:} \FunctionTok{\{}\DataTypeTok{"tool"}\FunctionTok{:} \FunctionTok{\{}\DataTypeTok{"type"}\FunctionTok{:} \StringTok{"string"}\FunctionTok{\},} \DataTypeTok{"relationship"}\FunctionTok{:} \FunctionTok{\{}\DataTypeTok{"type"}\FunctionTok{:} \StringTok{"string"}\FunctionTok{,} \DataTypeTok{"enum"}\FunctionTok{:}
            \OtherTok{[}\StringTok{"uses"}\OtherTok{,} \StringTok{"extends"}\OtherTok{,} \StringTok{"wraps"}\OtherTok{,} \StringTok{"compares\_against"}\OtherTok{,} \StringTok{"replaces"}\OtherTok{,} \StringTok{"integrates\_with"}\OtherTok{,}
             \StringTok{"visualizes"}\OtherTok{,} \StringTok{"depends\_on"}\OtherTok{]}\FunctionTok{\}\},}
        \DataTypeTok{"required"}\FunctionTok{:} \OtherTok{[}\StringTok{"tool"}\OtherTok{,} \StringTok{"relationship"}\OtherTok{]}\FunctionTok{\}\},}
    \DataTypeTok{"databases\_used"}\FunctionTok{:} \FunctionTok{\{}\DataTypeTok{"type"}\FunctionTok{:} \StringTok{"array"}\FunctionTok{,} \DataTypeTok{"items"}\FunctionTok{:} \FunctionTok{\{}\DataTypeTok{"type"}\FunctionTok{:} \StringTok{"string"}\FunctionTok{\}\},}
    \DataTypeTok{"summary"}\FunctionTok{:} \FunctionTok{\{}\DataTypeTok{"type"}\FunctionTok{:} \StringTok{"string"}\FunctionTok{\}}
  \FunctionTok{\},}
  \DataTypeTok{"required"}\FunctionTok{:} \OtherTok{[}\StringTok{"artifact\_type"}\OtherTok{,} \StringTok{"artifact\_name"}\OtherTok{,} \StringTok{"biological\_questions"}\OtherTok{,} \StringTok{"tool\_relationships"}\OtherTok{,}
               \StringTok{"databases\_used"}\OtherTok{,} \StringTok{"summary"}\OtherTok{]}
\FunctionTok{\}}
\end{Highlighting}
\end{Shaded}

\textbf{Table S3.} Composition of the clustering by parent topic, with the share of each parent's papers
that lie in clusters judged biomedical by Qwen3-8B and whether the parent was retained, which it is
when that share exceeds one half (\texttt{Table\_S3\_parent\_topics.csv})

\begin{longtable}[]{@{}
  >{\raggedright\arraybackslash}p{(\linewidth - 8\tabcolsep) * \real{0.4375}}
  >{\raggedleft\arraybackslash}p{(\linewidth - 8\tabcolsep) * \real{0.0804}}
  >{\raggedleft\arraybackslash}p{(\linewidth - 8\tabcolsep) * \real{0.0714}}
  >{\raggedleft\arraybackslash}p{(\linewidth - 8\tabcolsep) * \real{0.3750}}
  >{\centering\arraybackslash}p{(\linewidth - 8\tabcolsep) * \real{0.0357}}@{}}
\toprule\noalign{}
\begin{minipage}[b]{\linewidth}\raggedright
Parent topic
\end{minipage} & \begin{minipage}[b]{\linewidth}\raggedleft
Clusters
\end{minipage} & \begin{minipage}[b]{\linewidth}\raggedleft
Papers
\end{minipage} & \begin{minipage}[b]{\linewidth}\raggedleft
\% of papers in clusters judged biomedical
\end{minipage} & \begin{minipage}[b]{\linewidth}\centering
Retained
\end{minipage} \\
\midrule\noalign{}
\endhead
\bottomrule\noalign{}
\endlastfoot
Biomedical Image and Signal Processing & 529 & 80,989 & 73.4 & yes \\
Biological Data Analysis & 430 & 78,609 & 99.5 & yes \\
Computational Materials Science & 205 & 28,889 & 36.4 & no \\
Optimization and Decision-Making & 81 & 11,058 & 12.0 & no \\
Wireless Sensor Networks and Related Technologies & 66 & 8,918 & 7.9 & no \\
Quantum and Computational Physics & 45 & 8,613 & 3.6 & no \\
Covid-19 Computational Epidemiology & 32 & 5,146 & 99.2 & yes \\
IoT Security and Privacy & 13 & 2,370 & 36.1 & no \\
Microgrid Control and Optimization & 16 & 1,810 & 0.0 & no \\
CMOS and Nanoscale Electronics & 12 & 681 & 6.0 & no \\
Medical Virtual Reality & 8 & 652 & 100.0 & yes \\
Healthcare Geospatial Analysis & 1 & 36 & 100.0 & yes \\
\end{longtable}

\textbf{Table S4.} The unmatched-name rate under alternative bases (\texttt{Table\_S4\_sensitivity.csv}). All rows
use the registry snapshots of Table S11. The generic filter removes names composed entirely of
platform, method and container words, and the name-form filter removes multi-token names with
neither a capital letter nor a digit. Unrestricted rows count every tool-type paper, clustered or not.
Inclusion rows place papers outside the clustering at the nearest of the 1,438 centroids in the
bge-base embedding, with no similarity floor, and keep those in a retained cluster. Named models are
the 60,295 model-type papers carrying a name. In no registry means matching none of bio.tools, PyPI, CRAN, Bioconductor and Bioconda. Restriction rows replace the parent-level rule by the
cluster-level judgements of Table S13 or apply the same majority rule at the second-best parent cut

\begin{longtable}[]{@{}
  >{\raggedright\arraybackslash}p{(\linewidth - 6\tabcolsep) * \real{0.6042}}
  >{\raggedleft\arraybackslash}p{(\linewidth - 6\tabcolsep) * \real{0.1562}}
  >{\raggedleft\arraybackslash}p{(\linewidth - 6\tabcolsep) * \real{0.1562}}
  >{\raggedleft\arraybackslash}p{(\linewidth - 6\tabcolsep) * \real{0.0833}}@{}}
\toprule\noalign{}
\begin{minipage}[b]{\linewidth}\raggedright
Basis
\end{minipage} & \begin{minipage}[b]{\linewidth}\raggedleft
Distinct names
\end{minipage} & \begin{minipage}[b]{\linewidth}\raggedleft
In no registry
\end{minipage} & \begin{minipage}[b]{\linewidth}\raggedleft
\%
\end{minipage} \\
\midrule\noalign{}
\endhead
\bottomrule\noalign{}
\endlastfoot
All clusters, generic filter only & 199,793 & 161,838 & 81.0 \\
Retained parents, generic filter only & 136,700 & 108,407 & 79.3 \\
All clusters, generic and name-form filters & 177,254 & 139,566 & 78.7 \\
Retained parents, generic and name-form filters (reported) & 121,844 & 93,762 & 77.0 \\
Reported, with unassigned papers at their nearest centroid & 139,768 & 106,741 & 76.4 \\
Reported, with named models at their nearest centroid & 161,519 & 128,813 & 79.8 \\
Reported, with both & 179,283 & 141,735 & 79.1 \\
Clusters judged biomedical individually & 113,284 & 86,000 & 75.9 \\
Clusters retained by both rules & 105,461 & 79,222 & 75.1 \\
Second-best parent cut (K = 30), same majority rule & 106,855 & 80,295 & 75.1 \\
\end{longtable}

What the filters remove from the retained atlas (\texttt{name\_filters.py}). Distinct names are counted after
each combination of filters, and the examples are 15 strings drawn at random (seed 20260914) from those
removed by each filter

\begin{longtable}[]{@{}lrr@{}}
\toprule\noalign{}
Name filters applied & Distinct names & In no registry \% \\
\midrule\noalign{}
\endhead
\bottomrule\noalign{}
\endlastfoot
no filter & 137,674 & 79.4 \\
generic filter only & 136,700 & 79.3 \\
name-form filter only & 122,490 & 77.0 \\
both (reported) & 121,844 & 77.0 \\
\end{longtable}

\begin{longtable}[]{@{}
  >{\raggedright\arraybackslash}p{(\linewidth - 2\tabcolsep) * \real{0.4257}}
  >{\raggedright\arraybackslash}p{(\linewidth - 2\tabcolsep) * \real{0.5743}}@{}}
\toprule\noalign{}
\begin{minipage}[b]{\linewidth}\raggedright
Removed by the generic filter
\end{minipage} & \begin{minipage}[b]{\linewidth}\raggedright
Removed by the name-form filter
\end{minipage} \\
\midrule\noalign{}
\endhead
\bottomrule\noalign{}
\endlastfoot
R software & gene set bagging \\
new AI software & sylvan linear regression \\
decision tree method & toxic unit extrapolation method \\
SAS-tools & new regression model for estimating readthrough propensity \\
genetic algorithm-Bayesian network approach & pseudo-likelihood solution \\
Monte Carlo-based simulation method & diametric ratio method \\
classifier software & novel algorithm for optimizing treatment strategies \\
new automated algorithm & phenotypic diagnostic algorithm \\
novel method & gamete binning \\
3-CNN model & attribute-based stochastic ensemble model \\
new analysis algorithm & integrated gradients method \\
Network-based computational approach & robust t-statistic method \\
new bioinformatics framework & spectral area extraction method \\
new Monte Carlo method & custom image processing pipeline \\
deep ensemble learning method & hybrid-alignment annotation method \\
\end{longtable}

\textbf{Table S5.} All 1,000 retained clusters (\texttt{Table\_S5\_clusters.csv}). Columns: cluster identifier,
generated label, size, parent topic, the cluster-level biomedical judgement, top class-based TF-IDF
keywords, and nearest EDAM term by embedding similarity with its similarity. The EDAM columns are
descriptive, since arbitrary bioinformatics text reaches similar values (Table S13). The parent topics
are a navigational hierarchy

\textbf{Table S6.} Model-based extraction audit (\texttt{Table\_S6\_extraction\_audit.csv}, 300 rows). Columns: the
extracted artifact type and name, the verdict of the judge model, Qwen2.5-32B-Instruct, on each field,
and the reason it gave. The judge saw the title and abstract only. The judge agreed with the artifact type
for 253 of 300 records (0.843) and with the name for 188 of the 198 carrying a name (0.949), rates of
agreement between models and not of accuracy

\textbf{Table S7.} Licence resolution. An earlier pass left 50,506 of the 1,074,191 papers with an
unresolved licence. Each has been re-read from its own \texttt{\textless{}permissions\textgreater{}} block, including the
\texttt{xlink:href} and \texttt{license-type} attributes, and assigned by the rules below. A class is granted only
where an explicit Creative Commons licence says so. Nothing is inferred from silence, and a
free-to-read publisher programme is not a reuse licence. The assignment was checked against PMC's
own licence field (\texttt{oa.fcgi}) for a stratified sample of 36 papers and agrees in 36 of 36.
Per-paper assignments are in \texttt{Table\_S7\_licence\_resolution.csv}, and the rules and
their evidence in \texttt{Table\_S7\_licence\_rules.csv}

\begin{longtable}[]{@{}
  >{\raggedright\arraybackslash}p{(\linewidth - 6\tabcolsep) * \real{0.2308}}
  >{\raggedleft\arraybackslash}p{(\linewidth - 6\tabcolsep) * \real{0.3077}}
  >{\raggedright\arraybackslash}p{(\linewidth - 6\tabcolsep) * \real{0.2308}}
  >{\raggedright\arraybackslash}p{(\linewidth - 6\tabcolsep) * \real{0.2308}}@{}}
\toprule\noalign{}
\begin{minipage}[b]{\linewidth}\raggedright
Rule
\end{minipage} & \begin{minipage}[b]{\linewidth}\raggedleft
Papers
\end{minipage} & \begin{minipage}[b]{\linewidth}\raggedright
Class
\end{minipage} & \begin{minipage}[b]{\linewidth}\raggedright
Basis
\end{minipage} \\
\midrule\noalign{}
\endhead
\bottomrule\noalign{}
\endlastfoot
no-grant & 20,518 & restricted & Permissions block asserts copyright and grants nothing \\
elsevier-covid-no-licence & 10,912 & restricted & Elsevier COVID-19 resource centre, free to read with no reuse licence, PMC records \texttt{license="none"} \\
cc-by & 8,075 & commercial & Creative Commons Attribution, by URL or \texttt{license-type} \\
no-permissions-block & 3,367 & restricted & No permissions element in the source XML \\
publisher-research-only & 3,225 & restricted & Publisher terms permit reading and academic research only, PMC records \texttt{license="none"} \\
cc-by-nc-nd & 1,799 & noncommercial & CC BY-NC-ND \\
cc-by-nc & 1,253 & noncommercial & CC BY-NC \\
us-gov & 702 & restricted & US Government work, but PMC records \texttt{license="none"}, so no PMC redistribution grant \\
all-rights-reserved & 371 & restricted & Explicit reservation of all rights \\
cc-by-nc-sa & 216 & noncommercial & CC BY-NC-SA \\
cc-by-nd & 41 & commercial & CC BY-ND permits verbatim commercial redistribution \\
public-domain & 26 & restricted & Public-domain assertion without a PMC licence grant \\
subscription & 1 & restricted & Subscription content, no reuse granted \\
\textbf{Total} & \textbf{50,506} & & \\
\end{longtable}

Corpus totals after this pass are commercial 831,601 (77.42\%), noncommercial 203,470 (18.94\%) and
restricted 39,120 (3.64\%). There is no undetermined class.

\textbf{Table S8.} What is distributed, namely the passage-level index and the structured companion
store.

The passage-level index. The upper table gives chunk composition by section kind. A profile chunk
carrying title, abstract and extracted fields is generated for every paper, so every paper is
retrievable even before its body text is indexed. The lower table gives the cost of vector
quantization, measured as recall@100 against unquantized retrieval over an identical graph, using 40
probe queries against a 120,000-vector sample. Resident sizes are for the full 30.9 million chunks

\begin{longtable}[]{@{}
  >{\raggedright\arraybackslash}p{(\linewidth - 4\tabcolsep) * \real{0.3000}}
  >{\raggedleft\arraybackslash}p{(\linewidth - 4\tabcolsep) * \real{0.4000}}
  >{\raggedright\arraybackslash}p{(\linewidth - 4\tabcolsep) * \real{0.3000}}@{}}
\toprule\noalign{}
\begin{minipage}[b]{\linewidth}\raggedright
Section kind
\end{minipage} & \begin{minipage}[b]{\linewidth}\raggedleft
Passages
\end{minipage} & \begin{minipage}[b]{\linewidth}\raggedright
Indexed
\end{minipage} \\
\midrule\noalign{}
\endhead
\bottomrule\noalign{}
\endlastfoot
Profile (one per paper) & 1,074,191 & yes \\
Methods & 7,309,733 & yes \\
Results & 7,004,105 & yes \\
Figure and table captions & 1,784,017 & yes \\
Data availability & 127,168 & yes \\
Introduction, Discussion, Conclusion, unclassified & 13,614,243 & yes \\
\textbf{Total} & \textbf{30,913,457} & \textbf{30,913,457 (100\%)} \\
\end{longtable}

\begin{longtable}[]{@{}
  >{\raggedright\arraybackslash}p{(\linewidth - 6\tabcolsep) * \real{0.2143}}
  >{\raggedleft\arraybackslash}p{(\linewidth - 6\tabcolsep) * \real{0.2857}}
  >{\raggedleft\arraybackslash}p{(\linewidth - 6\tabcolsep) * \real{0.2857}}
  >{\raggedright\arraybackslash}p{(\linewidth - 6\tabcolsep) * \real{0.2143}}@{}}
\toprule\noalign{}
\begin{minipage}[b]{\linewidth}\raggedright
Quantization
\end{minipage} & \begin{minipage}[b]{\linewidth}\raggedleft
Recall@100 vs full precision
\end{minipage} & \begin{minipage}[b]{\linewidth}\raggedleft
Resident vectors
\end{minipage} & \begin{minipage}[b]{\linewidth}\raggedright
Gate ($\geq$98\%)
\end{minipage} \\
\midrule\noalign{}
\endhead
\bottomrule\noalign{}
\endlastfoot
Binary (in use) & 98.80\% & \textasciitilde4 GB & pass \\
int8 & 99.55\% & \textasciitilde31 GB & pass \\
\end{longtable}

The structured companion store. The internal citation graph is resolved through PubMed
and DOI identifiers, because JATS reference lists carry those and almost never a PMC identifier

\begin{longtable}[]{@{}
  >{\raggedright\arraybackslash}p{(\linewidth - 2\tabcolsep) * \real{0.4286}}
  >{\raggedleft\arraybackslash}p{(\linewidth - 2\tabcolsep) * \real{0.5714}}@{}}
\toprule\noalign{}
\begin{minipage}[b]{\linewidth}\raggedright
Contents
\end{minipage} & \begin{minipage}[b]{\linewidth}\raggedleft
Rows
\end{minipage} \\
\midrule\noalign{}
\endhead
\bottomrule\noalign{}
\endlastfoot
Papers & 1,074,191 \\
Parsed reference entries & 41,890,329 \\
--- resolving to another paper in the corpus & 5,767,772 \\
Dataset-accession occurrences (285,615 distinct, over 184,701 papers) & 499,531 \\
Extracted tool relationships & 597,117 \\
Tool mentions & 596,496 \\
Databases used & 592,614 \\
Code-link occurrences (77,382 distinct Git repositories, over 97,331 papers) & 152,097 \\
\end{longtable}

\textbf{Table S9.} Screening-model and clustering-encoder selection. Models were compared against a
reference set of 41,928 abstracts labelled by a different model, so the figures measure agreement
with that reference and not accuracy against a human standard. The first four rows are the selection
comparison, run under identical conditions. The 1,000-abstract subset is 17.1\% positive against
19.7\% in the full set, so agreement should not be compared across rows of differing \emph{n} at the third
decimal. The final row re-measures the selected model on the whole reference set, giving Cohen's
$\kappa$ = 0.652, per-class F1 0.716 computational and 0.936 non-computational, with every output inside
the two permitted tokens. Encoders were compared under identical data splits and an identical
classification head, so the embedding is the only variable. Screening took 141 GPU-hours and
extraction 53, on one NVIDIA RTX PRO 6000 Blackwell GPU

\begin{longtable}[]{@{}
  >{\raggedright\arraybackslash}p{(\linewidth - 8\tabcolsep) * \real{0.1579}}
  >{\raggedleft\arraybackslash}p{(\linewidth - 8\tabcolsep) * \real{0.2105}}
  >{\raggedleft\arraybackslash}p{(\linewidth - 8\tabcolsep) * \real{0.2105}}
  >{\raggedleft\arraybackslash}p{(\linewidth - 8\tabcolsep) * \real{0.2105}}
  >{\raggedleft\arraybackslash}p{(\linewidth - 8\tabcolsep) * \real{0.2105}}@{}}
\toprule\noalign{}
\begin{minipage}[b]{\linewidth}\raggedright
Screening model
\end{minipage} & \begin{minipage}[b]{\linewidth}\raggedleft
Eval n
\end{minipage} & \begin{minipage}[b]{\linewidth}\raggedleft
Agreement
\end{minipage} & \begin{minipage}[b]{\linewidth}\raggedleft
Macro-F1
\end{minipage} & \begin{minipage}[b]{\linewidth}\raggedleft
Abstracts/s
\end{minipage} \\
\midrule\noalign{}
\endhead
\bottomrule\noalign{}
\endlastfoot
Qwen2.5-3B-Instruct & 41,928 & 0.840 & 0.630 & 147.3 \\
Qwen2.5-7B-Instruct & 1,000 & 0.879 & 0.702 & 64.5 \\
Qwen2.5-14B-Instruct & 1,000 & 0.896 & 0.754 & 33.1 \\
Qwen2.5-32B-Instruct (selected) & 1,000 & 0.903 & 0.782 & 13.4 \\
Qwen2.5-32B-Instruct, full reference set & 41,845 & 0.895 & 0.826 & 10.2 \\
\end{longtable}

\begin{longtable}[]{@{}lrrrr@{}}
\toprule\noalign{}
Clustering encoder & Dim & Test F1 & Test ROC-AUC & Encode (s) \\
\midrule\noalign{}
\endhead
\bottomrule\noalign{}
\endlastfoot
bge-large-en-v1.5 & 1024 & 0.7040 & 0.9045 & 134.5 \\
bge-base-en-v1.5 (selected) & 768 & 0.6994 & 0.8997 & 56.4 \\
e5-large-v2 & 1024 & 0.6956 & 0.8965 & 134.2 \\
\end{longtable}

\textbf{Table S10.} Artifact types across all 1,074,191 papers screened as computational, with the 18,251 recovered
failures in place, and the computational share of screened papers by publication year
(\texttt{Table\_S10\_share\_by\_year.csv} gives every year). Shards follow PMC identifier order, and the year
gives the trend directly

\begin{longtable}[]{@{}lrr@{}}
\toprule\noalign{}
Artifact type & Count & \% \\
\midrule\noalign{}
\endhead
\bottomrule\noalign{}
\endlastfoot
analysis only & 302,765 & 28.19 \\
model & 290,327 & 27.03 \\
new algorithm & 215,061 & 20.02 \\
database & 67,327 & 6.27 \\
pipeline & 64,560 & 6.01 \\
other & 47,074 & 4.38 \\
software package & 42,455 & 3.95 \\
review & 23,799 & 2.22 \\
benchmark & 10,395 & 0.97 \\
web server & 10,377 & 0.97 \\
no valid output after rerun & 51 & 0.005 \\
\end{longtable}

\begin{longtable}[]{@{}lrrr@{}}
\toprule\noalign{}
Publication year & Screened & Computational & \% \\
\midrule\noalign{}
\endhead
\bottomrule\noalign{}
\endlastfoot
2000 & 3,837 & 191 & 4.98 \\
2005 & 13,140 & 2,464 & 18.75 \\
2007 & 24,266 & 4,883 & 20.12 \\
2010 & 72,465 & 10,409 & 14.36 \\
2012 & 132,790 & 17,910 & 13.49 \\
2015 & 231,409 & 34,906 & 15.08 \\
2018 & 332,194 & 52,077 & 15.68 \\
2020 & 538,113 & 84,009 & 15.61 \\
2022 & 769,730 & 135,777 & 17.64 \\
2024 & 729,792 & 137,191 & 18.8 \\
2025 & 616,485 & 122,968 & 19.95 \\
\end{longtable}

\textbf{Table S11.} bio.tools as a reference set. The upper table identifies the registry snapshots, all
retrieved in September 2026. The bio.tools snapshot was retrieved from the public API and is archived as \texttt{biotools\_full\_20260914.json.gz}, with the SHA-256 of its uncompressed
JSON (a5bdca4941372cdf28eb479720e9559973640be1ee9397a4c3641959d4e95b4d). The PyPI, CRAN,
Bioconductor and Bioconda name lists are archived in \texttt{registries\_20260914/} with the SHA-256 of each
raw response.
An entry's defining publications are its Primary publications, or its untyped publications when it
has no Primary one, and 250 entries carry more than one Primary publication. Publications were
resolved to the corpus by PMCID, PMID or DOI, and remaining PMIDs and DOIs through the PMC ID
converter. Each entry is followed through one paper fixed before its outcome was known, the earliest
of its defining publications in the corpus by bio.tools publication date and then PMCID. Per-entry
results for the 12,207 entries of the software subset are in \texttt{Table\_S11\_biotools\_recall.csv}
(bio.tools identifier and name, tool types, defining basis, defining papers in the corpus, PMCID
followed, entries sharing that publication, stage reached, extracted type and name, strict,
identity-resolved and any-paper recovery, and the identity evidence)

\begin{longtable}[]{@{}
  >{\raggedright\arraybackslash}p{(\linewidth - 6\tabcolsep) * \real{0.1481}}
  >{\raggedright\arraybackslash}p{(\linewidth - 6\tabcolsep) * \real{0.5432}}
  >{\raggedleft\arraybackslash}p{(\linewidth - 6\tabcolsep) * \real{0.0988}}
  >{\raggedright\arraybackslash}p{(\linewidth - 6\tabcolsep) * \real{0.2099}}@{}}
\toprule\noalign{}
\begin{minipage}[b]{\linewidth}\raggedright
Registry
\end{minipage} & \begin{minipage}[b]{\linewidth}\raggedright
Source
\end{minipage} & \begin{minipage}[b]{\linewidth}\raggedleft
Names
\end{minipage} & \begin{minipage}[b]{\linewidth}\raggedright
SHA-256
\end{minipage} \\
\midrule\noalign{}
\endhead
\bottomrule\noalign{}
\endlastfoot
bio.tools & bio.tools/api/tool/?format=json & 34,232 & a5bdca4941372cdf\ldots{} \\
PyPI & pypi.org/simple/ & 891,394 & 01964386a0dcbd39\ldots{} \\
CRAN & cran.r-project.org/src/contrib/PACKAGES & 25,023 & e6e0ea74adf3e276\ldots{} \\
Bioconductor & bioconductor.org/packages/release/bioc/VIEWS & 2,418 & c8b67a4faf1a48bd\ldots{} \\
Bioconda & conda.anaconda.org/bioconda/channeldata.json & 12,718 & 561fc0e3dc41178e\ldots{} \\
\end{longtable}

\begin{longtable}[]{@{}
  >{\raggedright\arraybackslash}p{(\linewidth - 4\tabcolsep) * \real{0.7838}}
  >{\raggedleft\arraybackslash}p{(\linewidth - 4\tabcolsep) * \real{0.1081}}
  >{\raggedleft\arraybackslash}p{(\linewidth - 4\tabcolsep) * \real{0.1081}}@{}}
\toprule\noalign{}
\begin{minipage}[b]{\linewidth}\raggedright
Furthest point reached by an entry's defining publications
\end{minipage} & \begin{minipage}[b]{\linewidth}\raggedleft
Entries
\end{minipage} & \begin{minipage}[b]{\linewidth}\raggedleft
\%
\end{minipage} \\
\midrule\noalign{}
\endhead
\bottomrule\noalign{}
\endlastfoot
no publication & 2,909 & 8.5 \\
only non-defining types & 906 & 2.6 \\
not in PMC & 11,044 & 32.3 \\
in PMC, not in the open-access corpus & 2,738 & 8.0 \\
parsed, no title and abstract & 160 & 0.5 \\
screened & 16,475 & 48.1 \\
Total & 34,232 & 100.0 \\
\end{longtable}

\begin{longtable}[]{@{}
  >{\raggedright\arraybackslash}p{(\linewidth - 2\tabcolsep) * \real{0.9121}}
  >{\raggedleft\arraybackslash}p{(\linewidth - 2\tabcolsep) * \real{0.0879}}@{}}
\toprule\noalign{}
\begin{minipage}[b]{\linewidth}\raggedright
Reference subset
\end{minipage} & \begin{minipage}[b]{\linewidth}\raggedleft
Entries
\end{minipage} \\
\midrule\noalign{}
\endhead
\bottomrule\noalign{}
\endlastfoot
Defining publication in the screened corpus (all tool types) & 16,475 \\
of which a tool type other than Database portal or Ontology (principal denominator) & 12,207 \\
of which Database portal or Ontology only & 1,321 \\
of which no tool type recorded & 2,947 \\
Principal entries whose followed publication describes no other entry & 11,498 \\
Principal entries with a Primary publication & 3,349 \\
Entries with more than one defining paper in the corpus & 355 \\
\end{longtable}

\begin{longtable}[]{@{}lr@{}}
\toprule\noalign{}
Entries per defining publication & Publications \\
\midrule\noalign{}
\endhead
\bottomrule\noalign{}
\endlastfoot
1 & 16,072 \\
2 & 313 \\
3-5 & 41 \\
6-20 & 10 \\
\textgreater20 & 1 \\
\end{longtable}

Included entries compared with excluded entries, as percentages of entries carrying each tool type
(an entry may carry several)

\begin{longtable}[]{@{}lrr@{}}
\toprule\noalign{}
& Included & Excluded \\
\midrule\noalign{}
\endhead
\bottomrule\noalign{}
\endlastfoot
Entries & 16,475 & 17,757 \\
Median first publication year & 2020 & 2019 \\
Web application, \% & 26.4 & 19.0 \\
Command-line tool, \% & 25.8 & 35.8 \\
Library, \% & 15.4 & 24.1 \\
Database portal, \% & 10.7 & 4.7 \\
Desktop application, \% & 5.4 & 6.7 \\
(none), \% & 17.9 & 16.3 \\
\end{longtable}

Selection of the reference set. Entries by how far their defining publications reach, with the first
bio.tools publication year (from publication metadata, absent for many entries) and the share carrying
each tool type

\begin{longtable}[]{@{}
  >{\raggedright\arraybackslash}p{(\linewidth - 6\tabcolsep) * \real{0.3617}}
  >{\raggedleft\arraybackslash}p{(\linewidth - 6\tabcolsep) * \real{0.2660}}
  >{\raggedleft\arraybackslash}p{(\linewidth - 6\tabcolsep) * \real{0.2553}}
  >{\raggedleft\arraybackslash}p{(\linewidth - 6\tabcolsep) * \real{0.1170}}@{}}
\toprule\noalign{}
\begin{minipage}[b]{\linewidth}\raggedright
\end{minipage} & \begin{minipage}[b]{\linewidth}\raggedleft
Screened (reference set)
\end{minipage} & \begin{minipage}[b]{\linewidth}\raggedleft
In PMC, not open access
\end{minipage} & \begin{minipage}[b]{\linewidth}\raggedleft
Not in PMC
\end{minipage} \\
\midrule\noalign{}
\endhead
\bottomrule\noalign{}
\endlastfoot
Entries & 16,475 & 2,738 & 11,044 \\
Median first publication year & 2020 & 2019 & 2020 \\
First publication before 2011, \% & 5.3 & 11.1 & 8.1 \\
First publication 2011-2015, \% & 8.7 & 9.1 & 4.3 \\
First publication 2016-2018, \% & 9.4 & 8.7 & 3.6 \\
First publication 2019-2021, \% & 27.9 & 24.0 & 16.5 \\
First publication 2022 or later, \% & 20.0 & 9.4 & 8.8 \\
No publication year, \% & 28.6 & 37.7 & 58.8 \\
Web application, \% & 26.4 & 23.7 & 17.6 \\
Command-line tool, \% & 25.8 & 34.0 & 38.4 \\
Library, \% & 15.4 & 15.3 & 24.2 \\
Database portal, \% & 10.7 & 6.0 & 4.6 \\
\end{longtable}

Recall, with Wilson 95\% intervals. Rows above the sensitivity rows are for the software subset,
followed through the pre-specified paper

\begin{longtable}[]{@{}
  >{\raggedright\arraybackslash}p{(\linewidth - 6\tabcolsep) * \real{0.6091}}
  >{\raggedleft\arraybackslash}p{(\linewidth - 6\tabcolsep) * \real{0.0727}}
  >{\raggedleft\arraybackslash}p{(\linewidth - 6\tabcolsep) * \real{0.1545}}
  >{\raggedleft\arraybackslash}p{(\linewidth - 6\tabcolsep) * \real{0.1636}}@{}}
\toprule\noalign{}
\begin{minipage}[b]{\linewidth}\raggedright
Measure
\end{minipage} & \begin{minipage}[b]{\linewidth}\raggedleft
n
\end{minipage} & \begin{minipage}[b]{\linewidth}\raggedleft
Strict
\end{minipage} & \begin{minipage}[b]{\linewidth}\raggedleft
Identity-resolved
\end{minipage} \\
\midrule\noalign{}
\endhead
\bottomrule\noalign{}
\endlastfoot
Entries & 12,207 & 56.6 (55.7-57.5) & 62.1 (61.3-63.0) \\
Publications followed & 11,831 & 58.2 (57.3-59.0) & 62.9 (62.0-63.8) \\
Entries from single-entry publications & 11,498 & 58.7 (57.8-59.6) & 63.2 (62.3-64.0) \\
Entries with a Primary publication & 3,349 & 54.3 (52.6-55.9) & 60.3 (58.7-62.0) \\
Entries whose paper entered the atlas & 8,265 & 83.6 (82.8-84.4) & 91.8 (91.1-92.3) \\
All compatible types admitted, no clustering & 12,207 & 81.3 (80.6-82.0) & 88.9 (88.4-89.5) \\
Sensitivity: recovered by any defining paper & 12,207 & 56.9 (56.1-57.8) & 62.6 (61.7-63.5) \\
Sensitivity: all tool types, including Database portal and Ontology & 16,475 & 51.2 (50.5-52.0) & 56.1 (55.4-56.9) \\
All tool types, entries whose paper entered the atlas & 10,071 & 83.8 (83.1-84.5) & 91.8 (91.3-92.4) \\
\end{longtable}

\begin{longtable}[]{@{}
  >{\raggedright\arraybackslash}p{(\linewidth - 8\tabcolsep) * \real{0.4400}}
  >{\raggedleft\arraybackslash}p{(\linewidth - 8\tabcolsep) * \real{0.1067}}
  >{\raggedleft\arraybackslash}p{(\linewidth - 8\tabcolsep) * \real{0.1067}}
  >{\raggedleft\arraybackslash}p{(\linewidth - 8\tabcolsep) * \real{0.2400}}
  >{\raggedleft\arraybackslash}p{(\linewidth - 8\tabcolsep) * \real{0.1067}}@{}}
\toprule\noalign{}
\begin{minipage}[b]{\linewidth}\raggedright
Stage at which the entry was lost
\end{minipage} & \begin{minipage}[b]{\linewidth}\raggedleft
Strict
\end{minipage} & \begin{minipage}[b]{\linewidth}\raggedleft
\%
\end{minipage} & \begin{minipage}[b]{\linewidth}\raggedleft
Identity-resolved
\end{minipage} & \begin{minipage}[b]{\linewidth}\raggedleft
\%
\end{minipage} \\
\midrule\noalign{}
\endhead
\bottomrule\noalign{}
\endlastfoot
screened as experimental & 95 & 0.8 & 95 & 0.8 \\
extraction did not parse & 0 & 0.0 & 0 & 0.0 \\
artifact type not clustered & 2,019 & 16.5 & 2,019 & 16.5 \\
left unassigned by HDBSCAN & 1,731 & 14.2 & 1,731 & 14.2 \\
non-biomedical parent topic & 97 & 0.8 & 97 & 0.8 \\
no usable name & 174 & 1.4 & 174 & 1.4 \\
name extracted, not matched & 1,181 & 9.7 & 507 & 4.2 \\
recovered & 6,910 & 56.6 & 7,584 & 62.1 \\
\end{longtable}

\begin{longtable}[]{@{}lr@{}}
\toprule\noalign{}
Artifact type of papers lost as not clustered & Entries \\
\midrule\noalign{}
\endhead
\bottomrule\noalign{}
\endlastfoot
database & 884 \\
model & 663 \\
pipeline & 354 \\
benchmark & 64 \\
analysis only & 40 \\
review & 8 \\
other & 6 \\
\end{longtable}

Recall by stratum, for the 12,207 entries of the software subset, as percentages of the stratum. An
entry carrying several tool types is counted under each. The name is in the title or abstract when the
entry's name, its name without a parenthetical, or its parenthetical acronym occurs as a whole word in
the followed paper's title or abstract. A code link is a repository or package link detected in the
followed paper

\begin{longtable}[]{@{}
  >{\raggedright\arraybackslash}p{(\linewidth - 12\tabcolsep) * \real{0.2000}}
  >{\raggedleft\arraybackslash}p{(\linewidth - 12\tabcolsep) * \real{0.0593}}
  >{\raggedleft\arraybackslash}p{(\linewidth - 12\tabcolsep) * \real{0.1259}}
  >{\raggedleft\arraybackslash}p{(\linewidth - 12\tabcolsep) * \real{0.1333}}
  >{\raggedleft\arraybackslash}p{(\linewidth - 12\tabcolsep) * \real{0.1852}}
  >{\raggedleft\arraybackslash}p{(\linewidth - 12\tabcolsep) * \real{0.1259}}
  >{\raggedleft\arraybackslash}p{(\linewidth - 12\tabcolsep) * \real{0.1704}}@{}}
\toprule\noalign{}
\begin{minipage}[b]{\linewidth}\raggedright
Stratum
\end{minipage} & \begin{minipage}[b]{\linewidth}\raggedleft
Entries
\end{minipage} & \begin{minipage}[b]{\linewidth}\raggedleft
Strict
\end{minipage} & \begin{minipage}[b]{\linewidth}\raggedleft
Identity-resolved
\end{minipage} & \begin{minipage}[b]{\linewidth}\raggedleft
Lost: type not clustered
\end{minipage} & \begin{minipage}[b]{\linewidth}\raggedleft
Lost: unassigned
\end{minipage} & \begin{minipage}[b]{\linewidth}\raggedleft
Lost: name not matched
\end{minipage} \\
\midrule\noalign{}
\endhead
\bottomrule\noalign{}
\endlastfoot
Year: before 2011 & 734 & 48.5 (44.9-52.1) & 58.0 & 15.3 & 14.6 & 19.3 \\
Year: 2011-2015 & 1,170 & 54.7 (51.8-57.5) & 61.2 & 14.5 & 12.5 & 14.9 \\
Year: 2016-2018 & 1,460 & 60.3 (57.8-62.8) & 65.8 & 11.6 & 16.2 & 8.5 \\
Year: 2019-2021 & 3,136 & 54.8 (53.0-56.5) & 59.3 & 20.7 & 14.4 & 6.4 \\
Year: 2022 or later & 2,938 & 58.0 (56.2-59.7) & 61.1 & 21.6 & 14.2 & 3.9 \\
Year: no year & 2,769 & 58.3 (56.4-60.1) & 65.9 & 10.2 & 13.4 & 15.3 \\
Type: Web application & 4,357 & 49.5 (48.0-50.9) & 54.9 & 24.4 & 14.1 & 9.8 \\
Type: Command-line tool & 4,248 & 62.2 (60.7-63.6) & 68.1 & 11.8 & 12.3 & 10.7 \\
Type: Library & 2,544 & 60.8 (58.9-62.7) & 67.6 & 7.2 & 17.6 & 10.7 \\
Type: Desktop application & 891 & 61.1 (57.8-64.2) & 66.9 & 6.5 & 19.3 & 10.0 \\
Type: Script & 555 & 58.2 (54.1-62.2) & 62.2 & 21.3 & 11.0 & 5.4 \\
Type: Database portal & 463 & 9.9 (7.5-13.0) & 12.3 & 81.6 & 3.5 & 3.0 \\
Type: Workflow & 420 & 54.8 (50.0-59.5) & 58.3 & 31.2 & 8.6 & 4.5 \\
Type: Web service & 304 & 38.8 (33.5-44.4) & 43.8 & 24.3 & 20.7 & 14.1 \\
Type: Suite & 207 & 58.9 (52.1-65.4) & 66.2 & 7.7 & 15.5 & 12.6 \\
Name in title/abstract: yes & 10,883 & 61.6 (60.7-62.5) & 65.7 & 15.4 & 14.8 & 5.8 \\
Name in title/abstract: no & 1,316 & 15.4 (13.6-17.5) & 33.1 & 25.8 & 8.7 & 41.7 \\
Code link: yes & 4,680 & 59.9 (58.5-61.3) & 65.8 & 15.3 & 15.6 & 7.2 \\
Code link: no & 7,527 & 54.6 (53.4-55.7) & 59.9 & 17.3 & 13.3 & 11.2 \\
\end{longtable}

\textbf{Table S12.} Scope crosswalk between the extracted artifact types and bio.tools. The first table
states, from the bio.tools curation guidelines and the extraction schema, whether each artifact type is
ordinarily registrable and under which tool types. The second gives registry presence by artifact type
as percentages of distinct names, strict alias matching adding to exact matching. New algorithms are
split by whether a code repository or package link was detected in the paper, and the named subset is
those whose linked repository or package name agrees with the extracted name. The third gives the
unmatched share with each registry left out in turn, the marginal contribution of that registry. The
fourth gives registry presence, by strict matching, for every artifact type the first table marks
registrable, over all topics and without clustering. The last gives, for bio.tools entries whose
defining paper was extracted, the share carrying each tool type (\texttt{Table\_S12\_type\_crosswalk.csv} gives
counts for all fifteen tool types). For identity resolution, names agree in form when one equals a comma-, slash-, colon-,
parenthesis- or ``and''-delimited part of the other, or contains the other with both at least five
characters, and the names of related entries are included; repository URLs listed by more than twenty
bio.tools entries are not counted as evidence

\begin{longtable}[]{@{}
  >{\raggedright\arraybackslash}p{(\linewidth - 4\tabcolsep) * \real{0.3333}}
  >{\raggedright\arraybackslash}p{(\linewidth - 4\tabcolsep) * \real{0.3333}}
  >{\raggedright\arraybackslash}p{(\linewidth - 4\tabcolsep) * \real{0.3333}}@{}}
\toprule\noalign{}
\begin{minipage}[b]{\linewidth}\raggedright
Extracted artifact type
\end{minipage} & \begin{minipage}[b]{\linewidth}\raggedright
Ordinarily registrable in bio.tools
\end{minipage} & \begin{minipage}[b]{\linewidth}\raggedright
Corresponding tool types
\end{minipage} \\
\midrule\noalign{}
\endhead
\bottomrule\noalign{}
\endlastfoot
software\_package & yes & Command-line tool, Library, Desktop application, Plug-in, Suite, Workbench \\
web\_server & yes & Web application, Web service, Web API, SPARQL endpoint \\
new\_algorithm & only when released as software & any software type \\
pipeline & yes, when released & Workflow, Command-line tool \\
database & yes, as a portal & Database portal \\
model & only when released as software & Web application, Command-line tool, Library, Script \\
benchmark & no, unless released as software & \\
analysis\_only, review, other & no & \\
\end{longtable}

\begin{longtable}[]{@{}
  >{\raggedright\arraybackslash}p{(\linewidth - 14\tabcolsep) * \real{0.3766}}
  >{\raggedleft\arraybackslash}p{(\linewidth - 14\tabcolsep) * \real{0.0519}}
  >{\raggedleft\arraybackslash}p{(\linewidth - 14\tabcolsep) * \real{0.1039}}
  >{\raggedleft\arraybackslash}p{(\linewidth - 14\tabcolsep) * \real{0.1039}}
  >{\raggedleft\arraybackslash}p{(\linewidth - 14\tabcolsep) * \real{0.1234}}
  >{\raggedleft\arraybackslash}p{(\linewidth - 14\tabcolsep) * \real{0.0584}}
  >{\raggedleft\arraybackslash}p{(\linewidth - 14\tabcolsep) * \real{0.0844}}
  >{\raggedleft\arraybackslash}p{(\linewidth - 14\tabcolsep) * \real{0.0974}}@{}}
\toprule\noalign{}
\begin{minipage}[b]{\linewidth}\raggedright
Names from
\end{minipage} & \begin{minipage}[b]{\linewidth}\raggedleft
Names
\end{minipage} & \begin{minipage}[b]{\linewidth}\raggedleft
bio.tools exact
\end{minipage} & \begin{minipage}[b]{\linewidth}\raggedleft
bio.tools alias
\end{minipage} & \begin{minipage}[b]{\linewidth}\raggedleft
bio.tools identity
\end{minipage} & \begin{minipage}[b]{\linewidth}\raggedleft
Packages
\end{minipage} & \begin{minipage}[b]{\linewidth}\raggedleft
None, strict
\end{minipage} & \begin{minipage}[b]{\linewidth}\raggedleft
None, identity
\end{minipage} \\
\midrule\noalign{}
\endhead
\bottomrule\noalign{}
\endlastfoot
all atlas names & 121,844 & 9.0 & 3.4 & 13.0 & 17.0 & 77.0 & 76.4 \\
software packages and web servers & 31,180 & 23.7 & 4.0 & 29.3 & 23.5 & 61.0 & 59.6 \\
software package & 25,241 & 22.2 & 3.7 & 27.2 & 25.6 & 61.7 & 60.6 \\
web server & 6,179 & 31.7 & 5.3 & 39.7 & 16.2 & 56.4 & 53.9 \\
new algorithm & 92,110 & 4.7 & 3.2 & 8.2 & 15.5 & 81.5 & 81.2 \\
new algorithm, code link detected & 14,491 & 12.7 & 3.9 & 17.7 & 23.7 & 69.3 & 68.3 \\
new algorithm, repository or package named as the artifact & 6,199 & 22.4 & 3.8 & 27.8 & 30.9 & 56.8 & 55.5 \\
new algorithm, no software evidence detected & 78,508 & 3.6 & 3.1 & 6.8 & 14.5 & 83.2 & 83.1 \\
\end{longtable}

\begin{longtable}[]{@{}
  >{\raggedright\arraybackslash}p{(\linewidth - 12\tabcolsep) * \real{0.2500}}
  >{\raggedleft\arraybackslash}p{(\linewidth - 12\tabcolsep) * \real{0.1288}}
  >{\raggedleft\arraybackslash}p{(\linewidth - 12\tabcolsep) * \real{0.1364}}
  >{\raggedleft\arraybackslash}p{(\linewidth - 12\tabcolsep) * \real{0.0985}}
  >{\raggedleft\arraybackslash}p{(\linewidth - 12\tabcolsep) * \real{0.0985}}
  >{\raggedleft\arraybackslash}p{(\linewidth - 12\tabcolsep) * \real{0.1591}}
  >{\raggedleft\arraybackslash}p{(\linewidth - 12\tabcolsep) * \real{0.1288}}@{}}
\toprule\noalign{}
\begin{minipage}[b]{\linewidth}\raggedright
Names from
\end{minipage} & \begin{minipage}[b]{\linewidth}\raggedleft
None of the five
\end{minipage} & \begin{minipage}[b]{\linewidth}\raggedleft
without bio.tools
\end{minipage} & \begin{minipage}[b]{\linewidth}\raggedleft
without PyPI
\end{minipage} & \begin{minipage}[b]{\linewidth}\raggedleft
without CRAN
\end{minipage} & \begin{minipage}[b]{\linewidth}\raggedleft
without Bioconductor
\end{minipage} & \begin{minipage}[b]{\linewidth}\raggedleft
without Bioconda
\end{minipage} \\
\midrule\noalign{}
\endhead
\bottomrule\noalign{}
\endlastfoot
all atlas names & 77.0 & 83.0 & 84.5 & 77.7 & 77.0 & 77.2 \\
software packages and web servers & 61.0 & 76.5 & 69.1 & 61.9 & 61.1 & 61.5 \\
\end{longtable}

\begin{longtable}[]{@{}
  >{\raggedright\arraybackslash}p{(\linewidth - 6\tabcolsep) * \real{0.5455}}
  >{\raggedleft\arraybackslash}p{(\linewidth - 6\tabcolsep) * \real{0.1039}}
  >{\raggedleft\arraybackslash}p{(\linewidth - 6\tabcolsep) * \real{0.1299}}
  >{\raggedleft\arraybackslash}p{(\linewidth - 6\tabcolsep) * \real{0.2208}}@{}}
\toprule\noalign{}
\begin{minipage}[b]{\linewidth}\raggedright
Names from papers extracted as, all topics
\end{minipage} & \begin{minipage}[b]{\linewidth}\raggedleft
Names
\end{minipage} & \begin{minipage}[b]{\linewidth}\raggedleft
bio.tools
\end{minipage} & \begin{minipage}[b]{\linewidth}\raggedleft
None of the five
\end{minipage} \\
\midrule\noalign{}
\endhead
\bottomrule\noalign{}
\endlastfoot
all compatible types & 254,585 & 9.5 & 81.1 \\
new algorithm & 134,085 & 6.6 & 83.0 \\
model & 46,135 & 5.1 & 86.9 \\
software packages and web servers & 45,105 & 24.7 & 63.7 \\
software package & 36,813 & 23.1 & 64.3 \\
database & 28,652 & 11.9 & 82.8 \\
web server & 8,695 & 33.5 & 59.2 \\
pipeline & 5,931 & 13.9 & 74.3 \\
\end{longtable}

Registry presence by the publication year of the earliest atlas paper giving each name, and by retained
parent topic (a name given by papers in several parents counts in each), by strict matching, for all
atlas names and for the core software subset. Every name is compared with the September 2026 snapshots,
so older names have had longer to be registered, and tools removed or renamed earlier count as
unmatched

\begin{longtable}[]{@{}
  >{\raggedright\arraybackslash}p{(\linewidth - 16\tabcolsep) * \real{0.2017}}
  >{\raggedleft\arraybackslash}p{(\linewidth - 16\tabcolsep) * \real{0.0672}}
  >{\raggedleft\arraybackslash}p{(\linewidth - 16\tabcolsep) * \real{0.1008}}
  >{\raggedleft\arraybackslash}p{(\linewidth - 16\tabcolsep) * \real{0.0924}}
  >{\raggedleft\arraybackslash}p{(\linewidth - 16\tabcolsep) * \real{0.0672}}
  >{\raggedleft\arraybackslash}p{(\linewidth - 16\tabcolsep) * \real{0.0924}}
  >{\raggedleft\arraybackslash}p{(\linewidth - 16\tabcolsep) * \real{0.1429}}
  >{\raggedleft\arraybackslash}p{(\linewidth - 16\tabcolsep) * \real{0.1345}}
  >{\raggedleft\arraybackslash}p{(\linewidth - 16\tabcolsep) * \real{0.1008}}@{}}
\toprule\noalign{}
\begin{minipage}[b]{\linewidth}\raggedright
Earliest paper naming it
\end{minipage} & \begin{minipage}[b]{\linewidth}\raggedleft
Names
\end{minipage} & \begin{minipage}[b]{\linewidth}\raggedleft
bio.tools \%
\end{minipage} & \begin{minipage}[b]{\linewidth}\raggedleft
Packages \%
\end{minipage} & \begin{minipage}[b]{\linewidth}\raggedleft
None \%
\end{minipage} & \begin{minipage}[b]{\linewidth}\raggedleft
Core names
\end{minipage} & \begin{minipage}[b]{\linewidth}\raggedleft
Core bio.tools \%
\end{minipage} & \begin{minipage}[b]{\linewidth}\raggedleft
Core packages \%
\end{minipage} & \begin{minipage}[b]{\linewidth}\raggedleft
Core none \%
\end{minipage} \\
\midrule\noalign{}
\endhead
\bottomrule\noalign{}
\endlastfoot
before 2011 & 7,891 & 18.8 & 20.8 & 69.6 & 2,810 & 35.9 & 25.8 & 53.5 \\
2011-2015 & 17,064 & 14.1 & 19.1 & 74.3 & 4,827 & 29.0 & 25.2 & 58.8 \\
2016-2018 & 15,561 & 12.7 & 18.6 & 75.9 & 4,222 & 28.7 & 24.6 & 60.1 \\
2019-2021 & 26,593 & 16.2 & 17.2 & 74.1 & 6,665 & 35.3 & 23.8 & 55.6 \\
2022 or later & 54,735 & 9.0 & 15.1 & 80.5 & 12,656 & 21.1 & 21.8 & 66.7 \\
\end{longtable}

\begin{longtable}[]{@{}
  >{\raggedright\arraybackslash}p{(\linewidth - 16\tabcolsep) * \real{0.2857}}
  >{\raggedleft\arraybackslash}p{(\linewidth - 16\tabcolsep) * \real{0.0602}}
  >{\raggedleft\arraybackslash}p{(\linewidth - 16\tabcolsep) * \real{0.0902}}
  >{\raggedleft\arraybackslash}p{(\linewidth - 16\tabcolsep) * \real{0.0827}}
  >{\raggedleft\arraybackslash}p{(\linewidth - 16\tabcolsep) * \real{0.0602}}
  >{\raggedleft\arraybackslash}p{(\linewidth - 16\tabcolsep) * \real{0.0827}}
  >{\raggedleft\arraybackslash}p{(\linewidth - 16\tabcolsep) * \real{0.1278}}
  >{\raggedleft\arraybackslash}p{(\linewidth - 16\tabcolsep) * \real{0.1203}}
  >{\raggedleft\arraybackslash}p{(\linewidth - 16\tabcolsep) * \real{0.0902}}@{}}
\toprule\noalign{}
\begin{minipage}[b]{\linewidth}\raggedright
Retained parent topic
\end{minipage} & \begin{minipage}[b]{\linewidth}\raggedleft
Names
\end{minipage} & \begin{minipage}[b]{\linewidth}\raggedleft
bio.tools \%
\end{minipage} & \begin{minipage}[b]{\linewidth}\raggedleft
Packages \%
\end{minipage} & \begin{minipage}[b]{\linewidth}\raggedleft
None \%
\end{minipage} & \begin{minipage}[b]{\linewidth}\raggedleft
Core names
\end{minipage} & \begin{minipage}[b]{\linewidth}\raggedleft
Core bio.tools \%
\end{minipage} & \begin{minipage}[b]{\linewidth}\raggedleft
Core packages \%
\end{minipage} & \begin{minipage}[b]{\linewidth}\raggedleft
Core none \%
\end{minipage} \\
\midrule\noalign{}
\endhead
\bottomrule\noalign{}
\endlastfoot
Biomedical Image and Signal Processing & 60,422 & 5.4 & 13.7 & 84.5 & 12,331 & 12.4 & 19.2 & 75.6 \\
Biological Data Analysis & 58,441 & 20.6 & 21.6 & 67.6 & 18,148 & 40.4 & 29.2 & 48.1 \\
Covid-19 Computational Epidemiology & 3,543 & 7.1 & 11.7 & 84.7 & 861 & 17.2 & 14.3 & 75.3 \\
Medical Virtual Reality & 507 & 4.1 & 10.8 & 87.8 & 463 & 3.7 & 10.4 & 88.1 \\
Healthcare Geospatial Analysis & 28 & 7.1 & 10.7 & 89.3 & 3 & 0.0 & 0.0 & 100.0 \\
\end{longtable}

\begin{longtable}[]{@{}
  >{\raggedright\arraybackslash}p{(\linewidth - 16\tabcolsep) * \real{0.1729}}
  >{\raggedleft\arraybackslash}p{(\linewidth - 16\tabcolsep) * \real{0.1128}}
  >{\raggedleft\arraybackslash}p{(\linewidth - 16\tabcolsep) * \real{0.1353}}
  >{\raggedleft\arraybackslash}p{(\linewidth - 16\tabcolsep) * \real{0.0602}}
  >{\raggedleft\arraybackslash}p{(\linewidth - 16\tabcolsep) * \real{0.1203}}
  >{\raggedleft\arraybackslash}p{(\linewidth - 16\tabcolsep) * \real{0.1203}}
  >{\raggedleft\arraybackslash}p{(\linewidth - 16\tabcolsep) * \real{0.0677}}
  >{\raggedleft\arraybackslash}p{(\linewidth - 16\tabcolsep) * \real{0.1504}}
  >{\raggedleft\arraybackslash}p{(\linewidth - 16\tabcolsep) * \real{0.0602}}@{}}
\toprule\noalign{}
\begin{minipage}[b]{\linewidth}\raggedright
Extracted artifact type
\end{minipage} & \begin{minipage}[b]{\linewidth}\raggedleft
Linked entries
\end{minipage} & \begin{minipage}[b]{\linewidth}\raggedleft
Command-line tool
\end{minipage} & \begin{minipage}[b]{\linewidth}\raggedleft
Library
\end{minipage} & \begin{minipage}[b]{\linewidth}\raggedleft
Web application
\end{minipage} & \begin{minipage}[b]{\linewidth}\raggedleft
Database portal
\end{minipage} & \begin{minipage}[b]{\linewidth}\raggedleft
Workflow
\end{minipage} & \begin{minipage}[b]{\linewidth}\raggedleft
Desktop application
\end{minipage} & \begin{minipage}[b]{\linewidth}\raggedleft
(none)
\end{minipage} \\
\midrule\noalign{}
\endhead
\bottomrule\noalign{}
\endlastfoot
software package & 6,236 & 32.7 & 25.5 & 14.4 & 1.0 & 3.0 & 10.8 & 18.6 \\
new algorithm & 3,573 & 43.6 & 20.1 & 10.0 & 0.5 & 2.4 & 3.0 & 19.1 \\
web server & 2,645 & 6.1 & 2.1 & 78.5 & 4.6 & 0.6 & 1.9 & 10.1 \\
database & 2,406 & 3.4 & 1.8 & 28.7 & 62.2 & 0.3 & 0.7 & 15.3 \\
model & 866 & 22.5 & 9.9 & 36.7 & 1.4 & 2.8 & 2.2 & 20.7 \\
pipeline & 490 & 36.5 & 7.1 & 9.6 & 1.4 & 19.6 & 2.4 & 25.7 \\
(screened out) & 216 & 7.4 & 6.9 & 24.1 & 19.0 & 0.5 & 4.2 & 42.1 \\
benchmark & 106 & 29.2 & 16.0 & 9.4 & 3.8 & 0.9 & 3.8 & 34.0 \\
analysis only & 79 & 22.8 & 11.4 & 15.2 & 7.6 & 1.3 & 3.8 & 39.2 \\
review & 23 & 17.4 & 13.0 & 0 & 4.3 & 4.3 & 4.3 & 60.9 \\
other & 10 & 40.0 & 0 & 10.0 & 10.0 & 0 & 20.0 & 30.0 \\
(parse failure) & 2 & 0 & 0 & 0 & 50.0 & 0 & 0 & 50.0 \\
\end{longtable}

\textbf{Table S13.} Validation of the categories. The reference partition was produced by \texttt{atlas/cluster.py}
with UMAP seeded and single-threaded, and HDBSCAN run on a separately cached seed-42 reduction
reproduces all 267,893 assignments exactly. Every stability run uses the same implementation
(umap-learn and scikit-learn) and changes one setting, and is compared with the reference assignment
(\texttt{Table\_S13\_cluster\_stability.csv}). A retained cluster persists when its best-matching cluster in the
run overlaps it at the stated Jaccard index. The last row repeats the whole pipeline without
the 54,205 papers lacking a recovered artifact name, and is compared with the reference assignment on the remaining papers (\texttt{unnamed\_ablation.py}). Runs at the five other UMAP seeds agree with one another at AMI 0.779-0.789. The threshold table places the manually set
reassignment threshold (0.80) and preprint threshold (0.78) against the similarity of papers to cluster
centroids. Parent topics were cut at the count maximizing centroid silhouette among 12 (0.118), 16
(0.093), 20 (0.093), 25 (0.101), 30 (0.104) and 40 (0.095). The EDAM embedding match uses the
operation and topic classes of a BioPortal EDAM export whose most recent terms date from EDAM 1.25,
797 non-obsolete terms each embedded as label and definition with bge-base, against cluster label and
eight keywords. Curated validation uses the identity-resolved single-publication bio.tools entries
linked to retained clusters, 1,000 permutations of cluster labels, and EDAM ancestors for the
embedding-term comparison; the curated test covers only the retained clusters with at least five linked entries. Cluster-level biomedical judgements were made by Qwen3-8B with thinking
disabled, from the cluster label, description, keywords and twelve representative titles, with no
unparsed answers (\texttt{Table\_S13\_cluster\_biomedical.csv}). The restriction table gives the retained
clusters, papers and unmatched-name rate under each rule, and for three rules the recall of the 12,207
bio.tools software entries by strict matching, end to end and among entries whose paper enters the atlas

\begin{longtable}[]{@{}
  >{\raggedright\arraybackslash}p{(\linewidth - 16\tabcolsep) * \real{0.3375}}
  >{\raggedleft\arraybackslash}p{(\linewidth - 16\tabcolsep) * \real{0.0562}}
  >{\raggedleft\arraybackslash}p{(\linewidth - 16\tabcolsep) * \real{0.0500}}
  >{\raggedleft\arraybackslash}p{(\linewidth - 16\tabcolsep) * \real{0.0500}}
  >{\raggedleft\arraybackslash}p{(\linewidth - 16\tabcolsep) * \real{0.0500}}
  >{\raggedleft\arraybackslash}p{(\linewidth - 16\tabcolsep) * \real{0.1438}}
  >{\raggedleft\arraybackslash}p{(\linewidth - 16\tabcolsep) * \real{0.1250}}
  >{\raggedleft\arraybackslash}p{(\linewidth - 16\tabcolsep) * \real{0.1250}}
  >{\raggedleft\arraybackslash}p{(\linewidth - 16\tabcolsep) * \real{0.0625}}@{}}
\toprule\noalign{}
\begin{minipage}[b]{\linewidth}\raggedright
Variant
\end{minipage} & \begin{minipage}[b]{\linewidth}\raggedleft
Clusters
\end{minipage} & \begin{minipage}[b]{\linewidth}\raggedleft
Noise \%
\end{minipage} & \begin{minipage}[b]{\linewidth}\raggedleft
AMI
\end{minipage} & \begin{minipage}[b]{\linewidth}\raggedleft
ARI
\end{minipage} & \begin{minipage}[b]{\linewidth}\raggedleft
AMI, clustered in both
\end{minipage} & \begin{minipage}[b]{\linewidth}\raggedleft
Median best Jaccard
\end{minipage} & \begin{minipage}[b]{\linewidth}\raggedleft
Persisting at J 0.5
\end{minipage} & \begin{minipage}[b]{\linewidth}\raggedleft
at J 0.75
\end{minipage} \\
\midrule\noalign{}
\endhead
\bottomrule\noalign{}
\endlastfoot
Reference (seed 42, 15 neighbours, 5 components, 20/5) & 1,438 & 15.0 & 1 & 1 & 1 & 1 & 1,000 & 1,000 \\
UMAP seed 0 & 1,459 & 15.4 & 0.7917 & 0.6169 & 0.8483 & 0.691 & 723 & 381 \\
UMAP seed 1 & 1,435 & 14.9 & 0.7825 & 0.5861 & 0.8408 & 0.658 & 679 & 350 \\
UMAP seed 2 & 1,483 & 15.0 & 0.7906 & 0.6167 & 0.8461 & 0.676 & 702 & 354 \\
UMAP seed 3 & 1,486 & 15.0 & 0.7876 & 0.6098 & 0.8442 & 0.661 & 699 & 332 \\
UMAP seed 4 & 1,474 & 15.2 & 0.7839 & 0.6035 & 0.8409 & 0.661 & 697 & 329 \\
30 UMAP neighbours & 1,198 & 17.4 & 0.7612 & 0.5638 & 0.8275 & 0.573 & 584 & 241 \\
10 UMAP components & 1,484 & 15.1 & 0.8117 & 0.6622 & 0.8595 & 0.729 & 759 & 457 \\
minimum cluster size 10 & 2,867 & 12.7 & 0.8246 & 0.675 & 0.8607 & 0.827 & 890 & 660 \\
minimum cluster size 40 & 747 & 17.5 & 0.8282 & 0.6651 & 0.871 & 0.313 & 459 & 358 \\
minimum samples 1 & 2,044 & 11.9 & 0.804 & 0.6308 & 0.8477 & 0.744 & 853 & 491 \\
minimum samples 10 & 1,149 & 17.1 & 0.8427 & 0.7176 & 0.8838 & 0.758 & 708 & 513 \\
Unnamed papers removed (213,688 papers) & 1,121 & 14.7 & 0.7391 & 0.4936 & 0.8084 & 0.546 & 547 & 216 \\
\end{longtable}

\begin{longtable}[]{@{}
  >{\raggedright\arraybackslash}p{(\linewidth - 12\tabcolsep) * \real{0.4217}}
  >{\raggedleft\arraybackslash}p{(\linewidth - 12\tabcolsep) * \real{0.0964}}
  >{\raggedleft\arraybackslash}p{(\linewidth - 12\tabcolsep) * \real{0.0964}}
  >{\raggedleft\arraybackslash}p{(\linewidth - 12\tabcolsep) * \real{0.0964}}
  >{\raggedleft\arraybackslash}p{(\linewidth - 12\tabcolsep) * \real{0.0964}}
  >{\raggedleft\arraybackslash}p{(\linewidth - 12\tabcolsep) * \real{0.0964}}
  >{\raggedleft\arraybackslash}p{(\linewidth - 12\tabcolsep) * \real{0.0964}}@{}}
\toprule\noalign{}
\begin{minipage}[b]{\linewidth}\raggedright
Cosine to centroid
\end{minipage} & \begin{minipage}[b]{\linewidth}\raggedleft
p5
\end{minipage} & \begin{minipage}[b]{\linewidth}\raggedleft
p10
\end{minipage} & \begin{minipage}[b]{\linewidth}\raggedleft
p25
\end{minipage} & \begin{minipage}[b]{\linewidth}\raggedleft
p50
\end{minipage} & \begin{minipage}[b]{\linewidth}\raggedleft
p75
\end{minipage} & \begin{minipage}[b]{\linewidth}\raggedleft
p90
\end{minipage} \\
\midrule\noalign{}
\endhead
\bottomrule\noalign{}
\endlastfoot
Clustered papers, own centroid & 0.784 & 0.804 & 0.822 & 0.843 & 0.863 & 0.88 \\
Unassigned papers, nearest centroid & 0.725 & 0.741 & 0.765 & 0.785 & 0.797 & 0.804 \\
\end{longtable}

Per-cluster stability, clustering validity and the reassignment threshold (\texttt{cluster\_quality.py}). A
cluster's stability is the median, over the five other UMAP seeds, of the Jaccard overlap with its
best-matching cluster, and every cluster's score is in \texttt{Table\_S13\_cluster\_stability\_scores.csv}. DBCV
(density-based clustering validation, Moulavi et al.~2014, from -1 to 1) was computed on 10,000 clustered papers drawn at random
(seed 20260914), in the UMAP space each partition was clustered in; it is negative for every partition,
consistent with the overlapping clusters shown by the silhouette. The threshold sweep reassigns the raw HDBSCAN noise of the reference run to the nearest raw centroid at each threshold, and reproduces the reference assignment exactly at 0.80. Similarities of clustered and unassigned papers to raw centroids
overlap too much to define a threshold from the data, so the sweep shows its consequences instead, including the recall of the 12,207 bio.tools software entries with the atlas rebuilt at each threshold (\texttt{threshold\_labels.py}, \texttt{biotools\_scope.py}). In the sweep, None \% and Core none \% are the unmatched shares of all and of core name strings, In atlas \% the share of entries whose paper enters the atlas, and Recall in atlas \% strict recall among those entries

Encoders, input texts and reductions compared on the clustering task (\texttt{repr\_compare.py}), on 15,000
tool-type papers comprising all 7,334 papers linked to a bio.tools entry with curated topics and 7,666
drawn at random (seed 20260914). Every configuration uses HDBSCAN at minimum cluster size 10 and
minimum samples 5 with excess-of-mass selection. Selected sections are those whose heading names
methods, implementation, software, algorithms, materials, or data or code availability, up to 2,000
characters, and 4,524 papers have none. DBCV is computed on the clustered papers in the space HDBSCAN
used. Text: TSQ, title, generated summary and biological questions (the atlas representation); TS, title and summary;
TA, title and abstract; TA+sections, title, abstract and selected sections. Reduction: UMAP to five components,
PCA to 50 or 100 components, or none (cosine on the unreduced embeddings). EDAM is topic concentration, the mean
share of linked papers carrying their cluster's most frequent curated topic over clusters with at least three
linked papers; Null is its mean over 200 label permutations; Tested is the number of such clusters. Seed AMI
compares partitions from two UMAP seeds. kNN is the share of each linked paper's ten nearest linked neighbors, by
cosine in the embedding, that share a curated topic, which depends on no clustering; random pairs share one
15.1-15.5\% of the time

\begin{longtable}[]{@{}
  >{\raggedright\arraybackslash}p{(\linewidth - 20\tabcolsep) * \real{0.1215}}
  >{\raggedright\arraybackslash}p{(\linewidth - 20\tabcolsep) * \real{0.1495}}
  >{\raggedright\arraybackslash}p{(\linewidth - 20\tabcolsep) * \real{0.1121}}
  >{\raggedleft\arraybackslash}p{(\linewidth - 20\tabcolsep) * \real{0.0841}}
  >{\raggedleft\arraybackslash}p{(\linewidth - 20\tabcolsep) * \real{0.0748}}
  >{\raggedleft\arraybackslash}p{(\linewidth - 20\tabcolsep) * \real{0.0748}}
  >{\raggedleft\arraybackslash}p{(\linewidth - 20\tabcolsep) * \real{0.0748}}
  >{\raggedleft\arraybackslash}p{(\linewidth - 20\tabcolsep) * \real{0.0748}}
  >{\raggedleft\arraybackslash}p{(\linewidth - 20\tabcolsep) * \real{0.0748}}
  >{\raggedleft\arraybackslash}p{(\linewidth - 20\tabcolsep) * \real{0.0841}}
  >{\raggedleft\arraybackslash}p{(\linewidth - 20\tabcolsep) * \real{0.0748}}@{}}
\toprule\noalign{}
\begin{minipage}[b]{\linewidth}\raggedright
Encoder
\end{minipage} & \begin{minipage}[b]{\linewidth}\raggedright
Text
\end{minipage} & \begin{minipage}[b]{\linewidth}\raggedright
Reduction
\end{minipage} & \begin{minipage}[b]{\linewidth}\raggedleft
Clusters
\end{minipage} & \begin{minipage}[b]{\linewidth}\raggedleft
Noise \%
\end{minipage} & \begin{minipage}[b]{\linewidth}\raggedleft
DBCV
\end{minipage} & \begin{minipage}[b]{\linewidth}\raggedleft
EDAM
\end{minipage} & \begin{minipage}[b]{\linewidth}\raggedleft
Null
\end{minipage} & \begin{minipage}[b]{\linewidth}\raggedleft
Tested
\end{minipage} & \begin{minipage}[b]{\linewidth}\raggedleft
Seed AMI
\end{minipage} & \begin{minipage}[b]{\linewidth}\raggedleft
kNN
\end{minipage} \\
\midrule\noalign{}
\endhead
\bottomrule\noalign{}
\endlastfoot
bge-base & TSQ (atlas) & UMAP & 252 & 38.8 & 0.286 & 0.636 & 0.243 & 154 & 0.676 & 0.582 \\
bge-base & TSQ (atlas) & PCA-50 & 2 & 44.7 & -0.112 & 0.142 & 0.122 & 1 & & 0.582 \\
bge-base & TSQ (atlas) & PCA-100 & 7 & 61.0 & -0.043 & 0.149 & 0.122 & 1 & & 0.582 \\
bge-base & TSQ (atlas) & none & 2 & 39.1 & 0.0 & 0.136 & 0.122 & 1 & & 0.582 \\
bge-base & TS & UMAP & 258 & 40.7 & 0.336 & 0.637 & 0.246 & 171 & 0.668 & 0.582 \\
bge-base & TA & UMAP & 284 & 33.4 & 0.38 & 0.645 & 0.247 & 198 & 0.714 & 0.629 \\
bge-base & TA & PCA-50 & 3 & 33.8 & -0.118 & 0.129 & 0.122 & 1 & & 0.629 \\
bge-base & TA & PCA-100 & 2 & 23.3 & -0.064 & 0.127 & 0.122 & 1 & & 0.629 \\
bge-base & TA & none & 5 & 51.8 & 0.037 & 0.567 & 0.263 & 2 & & 0.629 \\
bge-base & TA+sections & UMAP & 301 & 33.6 & 0.407 & 0.659 & 0.25 & 205 & 0.715 & 0.636 \\
bge-large & TSQ (atlas) & UMAP & 264 & 37.5 & 0.365 & 0.654 & 0.248 & 171 & 0.702 & 0.611 \\
bge-large & TA & UMAP & 295 & 33.2 & 0.417 & 0.652 & 0.249 & 219 & 0.721 & 0.642 \\
e5-large & TSQ (atlas) & UMAP & 231 & 46.0 & 0.281 & 0.606 & 0.243 & 151 & 0.638 & 0.556 \\
e5-large & TA & UMAP & 257 & 41.5 & 0.384 & 0.629 & 0.242 & 162 & 0.69 & 0.599 \\
\end{longtable}

\begin{longtable}[]{@{}
  >{\raggedright\arraybackslash}p{(\linewidth - 2\tabcolsep) * \real{0.7250}}
  >{\raggedleft\arraybackslash}p{(\linewidth - 2\tabcolsep) * \real{0.2750}}@{}}
\toprule\noalign{}
\begin{minipage}[b]{\linewidth}\raggedright
Per-cluster stability of the 1,000 retained clusters
\end{minipage} & \begin{minipage}[b]{\linewidth}\raggedleft
Value
\end{minipage} \\
\midrule\noalign{}
\endhead
\bottomrule\noalign{}
\endlastfoot
Median best-match Jaccard over five UMAP seeds, quartiles & 0.468 / 0.675 / 0.787 \\
Clusters with stability 0.5 or above, \% & 72.2 \\
Clusters with stability 0.75 or above, \% & 34.8 \\
Retained papers in clusters with stability 0.5 or above, \% & 76.0 \\
Rank correlation of stability with cluster size & 0.033 \\
\end{longtable}

\begin{longtable}[]{@{}lrr@{}}
\toprule\noalign{}
Partition & Clusters & DBCV \\
\midrule\noalign{}
\endhead
\bottomrule\noalign{}
\endlastfoot
Reference partition & 1,438 & -0.11 \\
minimum cluster size 10 & 2,867 & -0.009 \\
minimum cluster size 40 & 747 & -0.231 \\
minimum samples 1 & 2,044 & -0.043 \\
minimum samples 10 & 1,149 & -0.172 \\
30 UMAP neighbours & 1,198 & -0.161 \\
10 UMAP components & 1,484 & -0.143 \\
\end{longtable}

\begin{longtable}[]{@{}
  >{\raggedright\arraybackslash}p{(\linewidth - 14\tabcolsep) * \real{0.1293}}
  >{\raggedleft\arraybackslash}p{(\linewidth - 14\tabcolsep) * \real{0.1379}}
  >{\raggedleft\arraybackslash}p{(\linewidth - 14\tabcolsep) * \real{0.0690}}
  >{\raggedleft\arraybackslash}p{(\linewidth - 14\tabcolsep) * \real{0.1034}}
  >{\raggedleft\arraybackslash}p{(\linewidth - 14\tabcolsep) * \real{0.0948}}
  >{\raggedleft\arraybackslash}p{(\linewidth - 14\tabcolsep) * \real{0.1466}}
  >{\raggedleft\arraybackslash}p{(\linewidth - 14\tabcolsep) * \real{0.1638}}
  >{\raggedleft\arraybackslash}p{(\linewidth - 14\tabcolsep) * \real{0.1552}}@{}}
\toprule\noalign{}
\begin{minipage}[b]{\linewidth}\raggedright
Threshold
\end{minipage} & \begin{minipage}[b]{\linewidth}\raggedleft
Retained papers
\end{minipage} & \begin{minipage}[b]{\linewidth}\raggedleft
None \%
\end{minipage} & \begin{minipage}[b]{\linewidth}\raggedleft
Core none \%
\end{minipage} & \begin{minipage}[b]{\linewidth}\raggedleft
In atlas \%
\end{minipage} & \begin{minipage}[b]{\linewidth}\raggedleft
Recall, strict
\end{minipage} & \begin{minipage}[b]{\linewidth}\raggedleft
Recall, identity \%
\end{minipage} & \begin{minipage}[b]{\linewidth}\raggedleft
Recall in atlas \%
\end{minipage} \\
\midrule\noalign{}
\endhead
\bottomrule\noalign{}
\endlastfoot
0.76 & 185,735 & 76.58 & 60.82 & 78.3 & 65.8 (64.9-66.6) & 72.0 & 84.0 \\
0.78 & 178,566 & 76.73 & 60.81 & 74.6 & 62.5 (61.7-63.4) & 68.5 & 83.8 \\
0.80 (reported) & 165,432 & 76.95 & 61.04 & 67.7 & 56.6 (55.7-57.5) & 62.1 & 83.6 \\
0.82 & 145,759 & 77.09 & 61.23 & 58.9 & 49.0 (48.1-49.9) & 54.0 & 83.2 \\
0.84 & 122,665 & 77.16 & 61.4 & 49.6 & 41.0 (40.2-41.9) & 45.3 & 82.8 \\
none & 92,382 & 77.04 & 61.37 & 36.6 & 30.6 (29.8-31.4) & 33.5 & 83.6 \\
\end{longtable}

\begin{longtable}[]{@{}
  >{\raggedright\arraybackslash}p{(\linewidth - 8\tabcolsep) * \real{0.3721}}
  >{\raggedleft\arraybackslash}p{(\linewidth - 8\tabcolsep) * \real{0.0620}}
  >{\raggedleft\arraybackslash}p{(\linewidth - 8\tabcolsep) * \real{0.2248}}
  >{\raggedleft\arraybackslash}p{(\linewidth - 8\tabcolsep) * \real{0.1705}}
  >{\raggedleft\arraybackslash}p{(\linewidth - 8\tabcolsep) * \real{0.1705}}@{}}
\toprule\noalign{}
\begin{minipage}[b]{\linewidth}\raggedright
Cluster text
\end{minipage} & \begin{minipage}[b]{\linewidth}\raggedleft
n
\end{minipage} & \begin{minipage}[b]{\linewidth}\raggedleft
Median best-match similarity
\end{minipage} & \begin{minipage}[b]{\linewidth}\raggedleft
Share at 0.6 or above
\end{minipage} & \begin{minipage}[b]{\linewidth}\raggedleft
Share at 0.7 or above
\end{minipage} \\
\midrule\noalign{}
\endhead
\bottomrule\noalign{}
\endlastfoot
Retained clusters & 1,000 & 0.703 & 0.981 & 0.515 \\
Excluded clusters & 438 & 0.649 & 0.936 & 0.119 \\
Label with another cluster's keywords (20 draws) & 20,000 & 0.71 & 0.99 & 0.573 \\
Random label and keywords (20 draws) & 20,000 & 0.694 & 0.993 & 0.447 \\
\end{longtable}

\begin{longtable}[]{@{}
  >{\raggedright\arraybackslash}p{(\linewidth - 10\tabcolsep) * \real{0.4694}}
  >{\raggedleft\arraybackslash}p{(\linewidth - 10\tabcolsep) * \real{0.0918}}
  >{\raggedleft\arraybackslash}p{(\linewidth - 10\tabcolsep) * \real{0.0918}}
  >{\raggedleft\arraybackslash}p{(\linewidth - 10\tabcolsep) * \real{0.1735}}
  >{\raggedleft\arraybackslash}p{(\linewidth - 10\tabcolsep) * \real{0.0918}}
  >{\raggedleft\arraybackslash}p{(\linewidth - 10\tabcolsep) * \real{0.0816}}@{}}
\toprule\noalign{}
\begin{minipage}[b]{\linewidth}\raggedright
Curated annotation
\end{minipage} & \begin{minipage}[b]{\linewidth}\raggedleft
Clusters
\end{minipage} & \begin{minipage}[b]{\linewidth}\raggedleft
Observed
\end{minipage} & \begin{minipage}[b]{\linewidth}\raggedleft
Permutation null
\end{minipage} & \begin{minipage}[b]{\linewidth}\raggedleft
Null p95
\end{minipage} & \begin{minipage}[b]{\linewidth}\raggedleft
p
\end{minipage} \\
\midrule\noalign{}
\endhead
\bottomrule\noalign{}
\endlastfoot
Most frequent EDAM topic, share of entries & 296 & 0.617 & 0.264 & 0.27 & 0.001 \\
Most frequent EDAM operation, share of entries & 294 & 0.424 & 0.188 & 0.194 & 0.001 \\
Embedding-matched term among curated terms & 569 & 0.54 & 0.11 & 0.13 & \\
\end{longtable}

EDAM annotation concentration under alternative linkages, and its distribution over clusters. In each
permutation the cluster labels are shuffled and the most frequent topic of every cluster is recomputed,
and p = (1 + number of permutations reaching the observed value) / (1 + 1,000). The reported linkage uses
single-entry publications, so each publication contributes one entry. Over all entries, publications
describing several entries are also permuted as blocks, each publication keeping its entries together

\begin{longtable}[]{@{}
  >{\raggedright\arraybackslash}p{(\linewidth - 14\tabcolsep) * \real{0.2819}}
  >{\raggedleft\arraybackslash}p{(\linewidth - 14\tabcolsep) * \real{0.1007}}
  >{\raggedleft\arraybackslash}p{(\linewidth - 14\tabcolsep) * \real{0.1074}}
  >{\raggedleft\arraybackslash}p{(\linewidth - 14\tabcolsep) * \real{0.0604}}
  >{\raggedleft\arraybackslash}p{(\linewidth - 14\tabcolsep) * \real{0.1544}}
  >{\raggedleft\arraybackslash}p{(\linewidth - 14\tabcolsep) * \real{0.0537}}
  >{\raggedleft\arraybackslash}p{(\linewidth - 14\tabcolsep) * \real{0.1879}}
  >{\raggedleft\arraybackslash}p{(\linewidth - 14\tabcolsep) * \real{0.0537}}@{}}
\toprule\noalign{}
\begin{minipage}[b]{\linewidth}\raggedright
Linkage of entries to clusters
\end{minipage} & \begin{minipage}[b]{\linewidth}\raggedleft
Linked entries
\end{minipage} & \begin{minipage}[b]{\linewidth}\raggedleft
Clusters tested
\end{minipage} & \begin{minipage}[b]{\linewidth}\raggedleft
Observed
\end{minipage} & \begin{minipage}[b]{\linewidth}\raggedleft
Null, entries permuted
\end{minipage} & \begin{minipage}[b]{\linewidth}\raggedleft
p
\end{minipage} & \begin{minipage}[b]{\linewidth}\raggedleft
Null, publications permuted
\end{minipage} & \begin{minipage}[b]{\linewidth}\raggedleft
p
\end{minipage} \\
\midrule\noalign{}
\endhead
\bottomrule\noalign{}
\endlastfoot
strict, single-entry publications & 6,748 & 279 & 0.616 & 0.263 & 0.001 & same as entries & \\
identity-resolved, single-entry (reported) & 7,261 & 296 & 0.617 & 0.264 & 0.001 & same as entries & \\
identity-resolved, all entries & 7,584 & 298 & 0.617 & 0.261 & 0.001 & 0.264 & 0.001 \\
\end{longtable}

\begin{longtable}[]{@{}
  >{\raggedright\arraybackslash}p{(\linewidth - 6\tabcolsep) * \real{0.4580}}
  >{\raggedleft\arraybackslash}p{(\linewidth - 6\tabcolsep) * \real{0.0687}}
  >{\raggedleft\arraybackslash}p{(\linewidth - 6\tabcolsep) * \real{0.1908}}
  >{\raggedleft\arraybackslash}p{(\linewidth - 6\tabcolsep) * \real{0.2824}}@{}}
\toprule\noalign{}
\begin{minipage}[b]{\linewidth}\raggedright
Per-cluster share of entries carrying the most frequent term
\end{minipage} & \begin{minipage}[b]{\linewidth}\raggedleft
Clusters
\end{minipage} & \begin{minipage}[b]{\linewidth}\raggedleft
Quartiles (25 / 50 / 75)
\end{minipage} & \begin{minipage}[b]{\linewidth}\raggedleft
Permutation null, mean over clusters
\end{minipage} \\
\midrule\noalign{}
\endhead
\bottomrule\noalign{}
\endlastfoot
EDAM topic & 296 & 0.480 / 0.600 / 0.770 & 0.264 \\
EDAM operation & 294 & 0.286 / 0.400 / 0.535 & 0.188 \\
\end{longtable}

Retained clusters with and without enough linked entries for the curated test (\texttt{edam\_tested.py}), with the share of their papers by artifact type, with a recovered name, and in each parent topic

\begin{longtable}[]{@{}
  >{\raggedright\arraybackslash}p{(\linewidth - 4\tabcolsep) * \real{0.5258}}
  >{\raggedleft\arraybackslash}p{(\linewidth - 4\tabcolsep) * \real{0.3814}}
  >{\raggedleft\arraybackslash}p{(\linewidth - 4\tabcolsep) * \real{0.0928}}@{}}
\toprule\noalign{}
\begin{minipage}[b]{\linewidth}\raggedright
Retained clusters
\end{minipage} & \begin{minipage}[b]{\linewidth}\raggedleft
Tested (five or more linked entries)
\end{minipage} & \begin{minipage}[b]{\linewidth}\raggedleft
Untested
\end{minipage} \\
\midrule\noalign{}
\endhead
\bottomrule\noalign{}
\endlastfoot
Clusters & 296 & 704 \\
Papers & 78,674 & 86,758 \\
Median cluster size, papers & 179 & 92 \\
Median publication year & 2020 & 2022 \\
New algorithm, \% of papers & 71.3 & 85.0 \\
Software package, \% of papers & 22.3 & 12.5 \\
Web server, \% of papers & 6.4 & 2.5 \\
Name recovered, \% of papers & 82.8 & 73.8 \\
Biological Data Analysis, \% of papers & 85.1 & 25.3 \\
Biomedical Image and Signal Processing, \% of papers & 13.2 & 69.6 \\
Covid-19 Computational Epidemiology, \% of papers & 1.7 & 3.8 \\
\end{longtable}

\begin{longtable}[]{@{}llrr@{}}
\toprule\noalign{}
Parent-level rule & Cluster judged biomedical & Clusters & Papers \\
\midrule\noalign{}
\endhead
\bottomrule\noalign{}
\endlastfoot
retained & yes & 819 & 143,448 \\
retained & no & 181 & 21,984 \\
excluded & yes & 105 & 13,753 \\
excluded & no & 333 & 48,586 \\
\end{longtable}

\begin{longtable}[]{@{}
  >{\raggedright\arraybackslash}p{(\linewidth - 14\tabcolsep) * \real{0.3217}}
  >{\raggedleft\arraybackslash}p{(\linewidth - 14\tabcolsep) * \real{0.0783}}
  >{\raggedleft\arraybackslash}p{(\linewidth - 14\tabcolsep) * \real{0.0696}}
  >{\raggedleft\arraybackslash}p{(\linewidth - 14\tabcolsep) * \real{0.0696}}
  >{\raggedleft\arraybackslash}p{(\linewidth - 14\tabcolsep) * \real{0.0870}}
  >{\raggedleft\arraybackslash}p{(\linewidth - 14\tabcolsep) * \real{0.0696}}
  >{\raggedleft\arraybackslash}p{(\linewidth - 14\tabcolsep) * \real{0.1478}}
  >{\raggedleft\arraybackslash}p{(\linewidth - 14\tabcolsep) * \real{0.1565}}@{}}
\toprule\noalign{}
\begin{minipage}[b]{\linewidth}\raggedright
Restriction
\end{minipage} & \begin{minipage}[b]{\linewidth}\raggedleft
Clusters
\end{minipage} & \begin{minipage}[b]{\linewidth}\raggedleft
Papers
\end{minipage} & \begin{minipage}[b]{\linewidth}\raggedleft
Names
\end{minipage} & \begin{minipage}[b]{\linewidth}\raggedleft
Unmatched
\end{minipage} & \begin{minipage}[b]{\linewidth}\raggedleft
\%
\end{minipage} & \begin{minipage}[b]{\linewidth}\raggedleft
Recall, strict \%
\end{minipage} & \begin{minipage}[b]{\linewidth}\raggedleft
Recall in atlas \%
\end{minipage} \\
\midrule\noalign{}
\endhead
\bottomrule\noalign{}
\endlastfoot
Parent rule (reported) & 1,000 & 165,432 & 121,844 & 93,762 & 77.0 & 56.6 & 83.6 \\
Cluster classifications alone & 924 & 157,201 & 113,284 & 86,000 & 75.9 & 56.2 & 83.6 \\
Retained by both rules & 819 & 143,448 & 105,461 & 79,222 & 75.1 & & \\
Parent rule, second-best cut (K = 30) & 866 & 146,915 & 106,855 & 80,295 & 75.1 & & \\
No restriction & 1,438 & 227,771 & 152,845 & 120,995 & 79.2 & 57.3 & 83.6 \\
\end{longtable}

\textbf{Table S14.} Extraction reruns and duplicates. The 18,302 production failures comprise 15,687 records
with no output and 2,615 whose output reached the 512-token cap. All were rerun with the production model, prompt, schema
and input window and a 1,536-token output cap, and the records that parse replace the failures in every
analysis. For the window comparison, 1,000 papers drawn at random from the 1,055,889 production records
that parsed were extracted again at the production settings, which measures run-to-run variation, and
with the body allowed up to 29,768 tokens of the model's 32,768-token context and a 1,536-token output
cap (long context). The median long-context prompt was 7,751 tokens and 15 bodies still exceeded the
context.
Tool-type names in the sample pass the same generic and name-form filters as the atlas but are not
restricted to biomedical clusters. The duplicate table removes from the retained atlas every preprint
whose normalized title matches a record with a non-preprint DOI in the computational corpus

\begin{longtable}[]{@{}lr@{}}
\toprule\noalign{}
Failure rerun & Records \\
\midrule\noalign{}
\endhead
\bottomrule\noalign{}
\endlastfoot
Rerun & 18,302 \\
Now valid, included in every analysis & 18,251 \\
Still invalid & 51 \\
Of a tool type & 4,989 \\
analysis only & 5,402 \\
new algorithm & 3,944 \\
model & 3,764 \\
database & 1,672 \\
pipeline & 1,035 \\
other & 937 \\
software package & 788 \\
review & 279 \\
web server & 257 \\
benchmark & 173 \\
\end{longtable}

\begin{longtable}[]{@{}
  >{\raggedright\arraybackslash}p{(\linewidth - 12\tabcolsep) * \real{0.3525}}
  >{\raggedleft\arraybackslash}p{(\linewidth - 12\tabcolsep) * \real{0.0576}}
  >{\raggedleft\arraybackslash}p{(\linewidth - 12\tabcolsep) * \real{0.0863}}
  >{\raggedleft\arraybackslash}p{(\linewidth - 12\tabcolsep) * \real{0.0863}}
  >{\raggedleft\arraybackslash}p{(\linewidth - 12\tabcolsep) * \real{0.0863}}
  >{\raggedleft\arraybackslash}p{(\linewidth - 12\tabcolsep) * \real{0.1727}}
  >{\raggedleft\arraybackslash}p{(\linewidth - 12\tabcolsep) * \real{0.1583}}@{}}
\toprule\noalign{}
\begin{minipage}[b]{\linewidth}\raggedright
Runs compared
\end{minipage} & \begin{minipage}[b]{\linewidth}\raggedleft
Papers
\end{minipage} & \begin{minipage}[b]{\linewidth}\raggedleft
Same type \%
\end{minipage} & \begin{minipage}[b]{\linewidth}\raggedleft
Named \%
\end{minipage} & \begin{minipage}[b]{\linewidth}\raggedleft
Tool type \%
\end{minipage} & \begin{minipage}[b]{\linewidth}\raggedleft
Same name \%, both named
\end{minipage} & \begin{minipage}[b]{\linewidth}\raggedleft
Named in one run only
\end{minipage} \\
\midrule\noalign{}
\endhead
\bottomrule\noalign{}
\endlastfoot
Production against rerun at the production window & 1,000 & 97.3 & 30.3 / 30.2 & 25.5 / 25.5 & 98.3 & 4 / 3 \\
Production window against full article & 1,000 & 73.7 & 30.2 / 48.1 & 25.5 / 30.6 & 74.4 & 13 / 192 \\
As above, bodies longer than the window only & 954 & 72.4 & 30.1 / 48.8 & 25.4 / 30.7 & 73.0 & 13 / 192 \\
Production against full article & 1,000 & 73.1 & 30.3 / 48.1 & 25.5 / 30.6 & 74.0 & 14 / 192 \\
\end{longtable}

\begin{longtable}[]{@{}
  >{\raggedright\arraybackslash}p{(\linewidth - 6\tabcolsep) * \real{0.3077}}
  >{\raggedleft\arraybackslash}p{(\linewidth - 6\tabcolsep) * \real{0.3205}}
  >{\raggedleft\arraybackslash}p{(\linewidth - 6\tabcolsep) * \real{0.2692}}
  >{\raggedleft\arraybackslash}p{(\linewidth - 6\tabcolsep) * \real{0.1026}}@{}}
\toprule\noalign{}
\begin{minipage}[b]{\linewidth}\raggedright
Run
\end{minipage} & \begin{minipage}[b]{\linewidth}\raggedleft
Distinct tool-type names
\end{minipage} & \begin{minipage}[b]{\linewidth}\raggedleft
Matching no registry
\end{minipage} & \begin{minipage}[b]{\linewidth}\raggedleft
\%
\end{minipage} \\
\midrule\noalign{}
\endhead
\bottomrule\noalign{}
\endlastfoot
Production & 188 & 133 & 70.7 \\
Rerun, production window & 189 & 134 & 70.9 \\
Rerun, full article & 241 & 175 & 72.6 \\
\end{longtable}

\begin{longtable}[]{@{}
  >{\raggedright\arraybackslash}p{(\linewidth - 4\tabcolsep) * \real{0.3895}}
  >{\raggedleft\arraybackslash}p{(\linewidth - 4\tabcolsep) * \real{0.1579}}
  >{\raggedleft\arraybackslash}p{(\linewidth - 4\tabcolsep) * \real{0.4526}}@{}}
\toprule\noalign{}
\begin{minipage}[b]{\linewidth}\raggedright
Retained atlas
\end{minipage} & \begin{minipage}[b]{\linewidth}\raggedleft
With preprints
\end{minipage} & \begin{minipage}[b]{\linewidth}\raggedleft
Preprints with a published version removed
\end{minipage} \\
\midrule\noalign{}
\endhead
\bottomrule\noalign{}
\endlastfoot
Papers & 165,432 & 164,446 \\
Distinct names & 121,844 & 121,649 \\
Names matching no registry, \% & 76.95 & 76.94 \\
Software package and web server names & 31,180 & 31,120 \\
of which matching no registry, \% & 61.04 & 61.02 \\
\end{longtable}

\textbf{Table S15.} Validation of the structured store. Code links were normalized to one identity per Git
repository (host, owner and name, case-folded, without a trailing .git), package, SourceForge project or
Zenodo record, and a random sample of repositories and package pages was checked in September 2026
against the GitHub, GitLab and Bitbucket APIs and the current CRAN, Bioconductor and PyPI indexes
(\texttt{Table\_S15\_code\_link\_check.csv}). Resolving shows that a link is live, not that it implements the
reporting paper's artifact. Accessions were detected by 20 identifier patterns and by context-gated PDB
and UniProt patterns, and a distinct accession is a distinct pair of type and identifier. Case-folding
identifiers merges four further accessions, and dbGaP identifiers retain their version suffix, giving
1,396 versioned identifiers for 901 studies. The citation graph was checked by agreement between the
two join identifiers on all edges and, for 300 random edges, against Crossref reference lists
deposited by publishers and PubMed reference links (\texttt{Table\_S15\_citation\_check.csv}). Preprint
assignment is against the atlas reported here, at a manually set threshold of 0.78. The
deployed monthly job assigns preprints to an extended atlas that also contains named models, and its
counts differ

\begin{longtable}[]{@{}lrr@{}}
\toprule\noalign{}
Code link kind & Occurrences & Distinct identities \\
\midrule\noalign{}
\endhead
\bottomrule\noalign{}
\endlastfoot
repository & 133,748 & 77,382 \\
package page & 11,294 & 2,811 \\
project page & 4,032 & 1,276 \\
archive record & 2,812 & 2,501 \\
account or organisation page & 149 & 143 \\
not a repository & 62 & 0 \\
Total & 152,097 & \\
\end{longtable}

\begin{longtable}[]{@{}
  >{\raggedright\arraybackslash}p{(\linewidth - 8\tabcolsep) * \real{0.2388}}
  >{\raggedleft\arraybackslash}p{(\linewidth - 8\tabcolsep) * \real{0.1194}}
  >{\raggedleft\arraybackslash}p{(\linewidth - 8\tabcolsep) * \real{0.1343}}
  >{\raggedleft\arraybackslash}p{(\linewidth - 8\tabcolsep) * \real{0.1343}}
  >{\raggedleft\arraybackslash}p{(\linewidth - 8\tabcolsep) * \real{0.3731}}@{}}
\toprule\noalign{}
\begin{minipage}[b]{\linewidth}\raggedright
Sample
\end{minipage} & \begin{minipage}[b]{\linewidth}\raggedleft
n
\end{minipage} & \begin{minipage}[b]{\linewidth}\raggedleft
Resolves
\end{minipage} & \begin{minipage}[b]{\linewidth}\raggedleft
Archived
\end{minipage} & \begin{minipage}[b]{\linewidth}\raggedleft
Not found or not indexed
\end{minipage} \\
\midrule\noalign{}
\endhead
\bottomrule\noalign{}
\endlastfoot
Git repositories & 400 & 379 & 6 & 15 \\
Package pages & 100 & 84 & 0 & 16 \\
\end{longtable}

\begin{longtable}[]{@{}lrr@{}}
\toprule\noalign{}
Accession type & Occurrences & Distinct \\
\midrule\noalign{}
\endhead
\bottomrule\noalign{}
\endlastfoot
geo\_series & 149,681 & 43,876 \\
pdb & 113,292 & 56,004 \\
bioproject & 51,856 & 41,162 \\
sra\_run & 41,119 & 37,632 \\
clinicaltrial & 20,822 & 14,712 \\
geo\_sample & 14,991 & 13,165 \\
geo\_platform & 14,296 & 1,489 \\
uniprot & 13,568 & 9,963 \\
sra\_study & 12,352 & 9,867 \\
sra\_experiment & 11,709 & 10,679 \\
biosample & 11,642 & 11,280 \\
zenodo & 9,946 & 9,425 \\
All 22 types & 499,531 & 285,615 \\
\end{longtable}

\begin{longtable}[]{@{}
  >{\raggedright\arraybackslash}p{(\linewidth - 6\tabcolsep) * \real{0.6667}}
  >{\raggedleft\arraybackslash}p{(\linewidth - 6\tabcolsep) * \real{0.1190}}
  >{\raggedleft\arraybackslash}p{(\linewidth - 6\tabcolsep) * \real{0.1190}}
  >{\raggedleft\arraybackslash}p{(\linewidth - 6\tabcolsep) * \real{0.0952}}@{}}
\toprule\noalign{}
\begin{minipage}[b]{\linewidth}\raggedright
Citation-graph check
\end{minipage} & \begin{minipage}[b]{\linewidth}\raggedleft
Checked
\end{minipage} & \begin{minipage}[b]{\linewidth}\raggedleft
Agreeing
\end{minipage} & \begin{minipage}[b]{\linewidth}\raggedleft
\%
\end{minipage} \\
\midrule\noalign{}
\endhead
\bottomrule\noalign{}
\endlastfoot
DOI-joined edges whose reference also carries a PMID & 3,440,318 & 3,348,792 & 97.3 \\
PMID-joined edges whose reference also carries a DOI & 3,411,055 & 3,348,792 & 98.2 \\
Random edges listed by Crossref or PubMed reference data & 298 & 294 & 98.7 \\
\end{longtable}

Why most reference entries do not enter the citation graph (\texttt{store\_coverage.py}). An entry resolves when
its PMCID, PMID or DOI matches a paper in the corpus; the rest identify a paper outside the
1,074,191-paper computational corpus or carry only malformed identifiers. Every stored entry carries
at least one identifier. Edges are counted per join identifier, so the 5,767,772 edges of Table S8
slightly exceed the entries that resolve

\begin{longtable}[]{@{}
  >{\raggedright\arraybackslash}p{(\linewidth - 4\tabcolsep) * \real{0.7816}}
  >{\raggedleft\arraybackslash}p{(\linewidth - 4\tabcolsep) * \real{0.1264}}
  >{\raggedleft\arraybackslash}p{(\linewidth - 4\tabcolsep) * \real{0.0920}}@{}}
\toprule\noalign{}
\begin{minipage}[b]{\linewidth}\raggedright
Parsed reference entries
\end{minipage} & \begin{minipage}[b]{\linewidth}\raggedleft
Entries
\end{minipage} & \begin{minipage}[b]{\linewidth}\raggedleft
\%
\end{minipage} \\
\midrule\noalign{}
\endhead
\bottomrule\noalign{}
\endlastfoot
resolve to a paper in the corpus & 5,748,812 & 13.7 \\
carry a well-formed PMID, DOI or PMCID of a paper outside the corpus & 36,140,959 & 86.3 \\
carry only malformed identifiers & 558 & 0.0 \\
carry no identifier & 0 & 0.0 \\
All & 41,890,329 & 100.0 \\
\end{longtable}

\begin{longtable}[]{@{}
  >{\raggedright\arraybackslash}p{(\linewidth - 2\tabcolsep) * \real{0.8904}}
  >{\raggedleft\arraybackslash}p{(\linewidth - 2\tabcolsep) * \real{0.1096}}@{}}
\toprule\noalign{}
\begin{minipage}[b]{\linewidth}\raggedright
Duplicate check across 1,074,191 papers
\end{minipage} & \begin{minipage}[b]{\linewidth}\raggedleft
Count
\end{minipage} \\
\midrule\noalign{}
\endhead
\bottomrule\noalign{}
\endlastfoot
Records with a bioRxiv or medRxiv DOI & 13,163 \\
Records with a Research Square DOI & 1,461 \\
Groups of records sharing a normalized title under different DOIs & 3,892 \\
Records in those groups & 7,860 \\
Groups containing more than one tool-type record & 1,229 \\
Groups containing a preprint-server DOI & 3,316 \\
\end{longtable}

\begin{longtable}[]{@{}
  >{\raggedright\arraybackslash}p{(\linewidth - 12\tabcolsep) * \real{0.1084}}
  >{\raggedleft\arraybackslash}p{(\linewidth - 12\tabcolsep) * \real{0.1325}}
  >{\raggedleft\arraybackslash}p{(\linewidth - 12\tabcolsep) * \real{0.1205}}
  >{\raggedleft\arraybackslash}p{(\linewidth - 12\tabcolsep) * \real{0.1084}}
  >{\raggedleft\arraybackslash}p{(\linewidth - 12\tabcolsep) * \real{0.2289}}
  >{\raggedleft\arraybackslash}p{(\linewidth - 12\tabcolsep) * \real{0.2048}}
  >{\raggedleft\arraybackslash}p{(\linewidth - 12\tabcolsep) * \real{0.0964}}@{}}
\toprule\noalign{}
\begin{minipage}[b]{\linewidth}\raggedright
Preprints
\end{minipage} & \begin{minipage}[b]{\linewidth}\raggedleft
Classified
\end{minipage} & \begin{minipage}[b]{\linewidth}\raggedleft
Tool type
\end{minipage} & \begin{minipage}[b]{\linewidth}\raggedleft
Assigned
\end{minipage} & \begin{minipage}[b]{\linewidth}\raggedleft
Assigned, retained
\end{minipage} & \begin{minipage}[b]{\linewidth}\raggedleft
Held as emerging
\end{minipage} & \begin{minipage}[b]{\linewidth}\raggedleft
\%
\end{minipage} \\
\midrule\noalign{}
\endhead
\bottomrule\noalign{}
\endlastfoot
bioRxiv & 123,939 & 22,502 & 21,301 & 19,595 & 1,201 & 5.3 \\
arXiv & 12,235 & 6,127 & 5,886 & 5,266 & 241 & 3.9 \\
\end{longtable}

Share of tool-type preprints left unassigned (held as emerging) at other similarity thresholds, over
the same nearest-centroid similarities

\begin{longtable}[]{@{}
  >{\raggedright\arraybackslash}p{(\linewidth - 12\tabcolsep) * \real{0.4343}}
  >{\raggedleft\arraybackslash}p{(\linewidth - 12\tabcolsep) * \real{0.0808}}
  >{\raggedleft\arraybackslash}p{(\linewidth - 12\tabcolsep) * \real{0.0808}}
  >{\raggedleft\arraybackslash}p{(\linewidth - 12\tabcolsep) * \real{0.1616}}
  >{\raggedleft\arraybackslash}p{(\linewidth - 12\tabcolsep) * \real{0.0808}}
  >{\raggedleft\arraybackslash}p{(\linewidth - 12\tabcolsep) * \real{0.0808}}
  >{\raggedleft\arraybackslash}p{(\linewidth - 12\tabcolsep) * \real{0.0808}}@{}}
\toprule\noalign{}
\begin{minipage}[b]{\linewidth}\raggedright
Preprints held as emerging, \%, at threshold
\end{minipage} & \begin{minipage}[b]{\linewidth}\raggedleft
0.74
\end{minipage} & \begin{minipage}[b]{\linewidth}\raggedleft
0.76
\end{minipage} & \begin{minipage}[b]{\linewidth}\raggedleft
0.78 (deployed)
\end{minipage} & \begin{minipage}[b]{\linewidth}\raggedleft
0.80
\end{minipage} & \begin{minipage}[b]{\linewidth}\raggedleft
0.82
\end{minipage} & \begin{minipage}[b]{\linewidth}\raggedleft
0.84
\end{minipage} \\
\midrule\noalign{}
\endhead
\bottomrule\noalign{}
\endlastfoot
bioRxiv & 0.7 & 1.9 & 5.3 & 13.1 & 28.1 & 52.7 \\
arXiv & 0.6 & 1.7 & 3.9 & 10.1 & 24.6 & 49.5 \\
\end{longtable}

Known-item retrieval test of the passage index (\texttt{retrieval\_benchmark.py}). The queries are the curated
descriptions of 300 bio.tools software entries drawn at random (seed 20260914) from the 9,620 whose
followed defining paper was screened computational and whose description keeps at least 80 characters
once the tool's name, its form without a parenthetical and its acronym are masked. The one relevant
paper is the defining paper. Each mode retrieves 200 passages, collapsed to papers in rank order.
Percentages and MRR with 95\% bootstrap intervals (2,000 resamples). Curated descriptions often
paraphrase the abstract, so this is an easy test of whether the index finds a described tool and not a
measure of topical search

\begin{longtable}[]{@{}
  >{\raggedright\arraybackslash}p{(\linewidth - 8\tabcolsep) * \real{0.3426}}
  >{\raggedleft\arraybackslash}p{(\linewidth - 8\tabcolsep) * \real{0.1574}}
  >{\raggedleft\arraybackslash}p{(\linewidth - 8\tabcolsep) * \real{0.1574}}
  >{\raggedleft\arraybackslash}p{(\linewidth - 8\tabcolsep) * \real{0.1574}}
  >{\raggedleft\arraybackslash}p{(\linewidth - 8\tabcolsep) * \real{0.1852}}@{}}
\toprule\noalign{}
\begin{minipage}[b]{\linewidth}\raggedright
Retrieval mode
\end{minipage} & \begin{minipage}[b]{\linewidth}\raggedleft
Recall@1
\end{minipage} & \begin{minipage}[b]{\linewidth}\raggedleft
Recall@10
\end{minipage} & \begin{minipage}[b]{\linewidth}\raggedleft
Recall@100
\end{minipage} & \begin{minipage}[b]{\linewidth}\raggedleft
MRR@10
\end{minipage} \\
\midrule\noalign{}
\endhead
\bottomrule\noalign{}
\endlastfoot
BM25 & 79.7 (75.0-84.0) & 90.3 (86.7-93.7) & 95.0 (92.3-97.3) & 0.833 (0.793-0.870) \\
dense & 75.3 (70.7-80.0) & 90.3 (87.0-93.3) & 97.3 (95.3-99.0) & 0.808 (0.769-0.846) \\
BM25 + dense (RRF) & 82.7 (78.3-87.0) & 92.7 (89.7-95.7) & 97.7 (96.0-99.3) & 0.865 (0.829-0.900) \\
RRF + cross-encoder rerank (deployed) & 82.7 (78.3-87.0) & 93.7 (90.7-96.3) & 97.3 (95.3-99.0) & 0.864 (0.830-0.899) \\
\end{longtable}

\textbf{Table S16.} Human audit. One annotator labels, for four sheets (S, screening; E, extraction;
U, unmatched names; M, matched names), a random subsample of each
stratum drawn with a fixed seed by \texttt{annotate.py} (its PLAN), rows shown round-robin across strata so a
partial pass still samples every stratum. The tool shows each sheet's decision rule and hides any
model decision that would bias the label (the screen's class and the extracted artifact type), and
records the annotator's training and field in \texttt{annotators.json}. Every sheet carries its stratum and
the stratum's population, and \texttt{score\_annotations.py} reports stratified estimates, each stratum
weighted by its population, with 95\% intervals from 2,000 bootstrap resamples drawn within strata.
UNCLEAR is reported as its own share, excluded from the primary estimate, and bounded by counting it
wholly for and against. The extraction sheet samples every extracted artifact type, so it gives a type
confusion matrix and the share of tool papers extracted as another type. With one annotator there is no
measure of inter-annotator agreement. Seven further sheets (identity resolutions, window
disagreements, software evidence, repositories, accessions, cluster labels and restriction
disagreements) are released unlabelled.
Sheets are drawn by \texttt{human\_samples.py}, \texttt{biotools\_scope.py}, \texttt{restriction\_sensitivity.py} and
\texttt{store\_checks.py} with seed 20260914

\begin{longtable}[]{@{}
  >{\raggedright\arraybackslash}p{(\linewidth - 4\tabcolsep) * \real{0.0196}}
  >{\raggedright\arraybackslash}p{(\linewidth - 4\tabcolsep) * \real{0.6816}}
  >{\raggedright\arraybackslash}p{(\linewidth - 4\tabcolsep) * \real{0.2989}}@{}}
\toprule\noalign{}
\begin{minipage}[b]{\linewidth}\raggedright
Set
\end{minipage} & \begin{minipage}[b]{\linewidth}\raggedright
Strata: sampled of population
\end{minipage} & \begin{minipage}[b]{\linewidth}\raggedright
Decision
\end{minipage} \\
\midrule\noalign{}
\endhead
\bottomrule\noalign{}
\endlastfoot
S & n = 150; screen positive: 50 of 1,074,191; keyword negative: 50 of 305,404; no-keyword negative: 50 of 5,065,730 & Is computation a central scientific result, by the Table S1 criterion? (screen label hidden, rows shuffled) \\
E & n = 13; analysis only: 1 of 302,765; model: 1 of 290,327; algorithm, name recovered: 4 of 162,952; algorithm, no name recovered: 1 of 52,109; pipeline/database: 1 of 131,887; review/benchmark/other: 1 of 81,268; software/web server: 4 of 52,832 & Are the type and name correct against the full article, and where is the name first stated? \\
U & n = 15; code-linked algorithm: 5 of 9,966; no-evidence algorithm: 5 of 64,765; software/web server: 5 of 19,031 & Is the name a genuine named method or software? If so, is it registered under another name? \\
M & n = 50; longer name: 25 of 24,949; short name: 25 of 3,133 & Is the matched registry record the same artifact as the paper's? \\
\end{longtable}

Raw answer counts by stratum, before weighting, for the primary question of each sheet: S, is computation a central result; E, is the extracted name correct; U, is the name a genuine method or software; M, is the matched record the same artifact. The estimates in the main text weight these by stratum population

\begin{longtable}[]{@{}
  >{\raggedright\arraybackslash}p{(\linewidth - 10\tabcolsep) * \real{0.1029}}
  >{\raggedright\arraybackslash}p{(\linewidth - 10\tabcolsep) * \real{0.4118}}
  >{\raggedleft\arraybackslash}p{(\linewidth - 10\tabcolsep) * \real{0.1324}}
  >{\raggedleft\arraybackslash}p{(\linewidth - 10\tabcolsep) * \real{0.1176}}
  >{\raggedleft\arraybackslash}p{(\linewidth - 10\tabcolsep) * \real{0.1176}}
  >{\raggedleft\arraybackslash}p{(\linewidth - 10\tabcolsep) * \real{0.1176}}@{}}
\toprule\noalign{}
\begin{minipage}[b]{\linewidth}\raggedright
Set
\end{minipage} & \begin{minipage}[b]{\linewidth}\raggedright
Stratum
\end{minipage} & \begin{minipage}[b]{\linewidth}\raggedleft
Labelled
\end{minipage} & \begin{minipage}[b]{\linewidth}\raggedleft
Yes
\end{minipage} & \begin{minipage}[b]{\linewidth}\raggedleft
No
\end{minipage} & \begin{minipage}[b]{\linewidth}\raggedleft
Unclear
\end{minipage} \\
\midrule\noalign{}
\endhead
\bottomrule\noalign{}
\endlastfoot
S & screen positive & 50 & 37 & 13 & 0 \\
S & keyword negative & 50 & 10 & 40 & 0 \\
S & no-keyword negative & 50 & 0 & 50 & 0 \\
E & analysis only & 1 & 0 & 0 & 1 \\
E & model & 1 & 1 & 0 & 0 \\
E & algorithm, name recovered & 4 & 4 & 0 & 0 \\
E & algorithm, no name recovered & 1 & 1 & 0 & 0 \\
E & pipeline/database & 1 & 1 & 0 & 0 \\
E & review/benchmark/other & 1 & 1 & 0 & 0 \\
E & software/web server & 4 & 4 & 0 & 0 \\
U & code-linked algorithm & 5 & 5 & 0 & 0 \\
U & no-evidence algorithm & 5 & 5 & 0 & 0 \\
U & software/web server & 5 & 5 & 0 & 0 \\
M & longer name & 25 & 14 & 11 & 0 \\
M & short name & 25 & 12 & 13 & 0 \\
\end{longtable}

\subsection{Supplementary figures}\label{supplementary-figures}

\pandocbounded{\includegraphics[keepaspectratio,alt={Figure S1}]{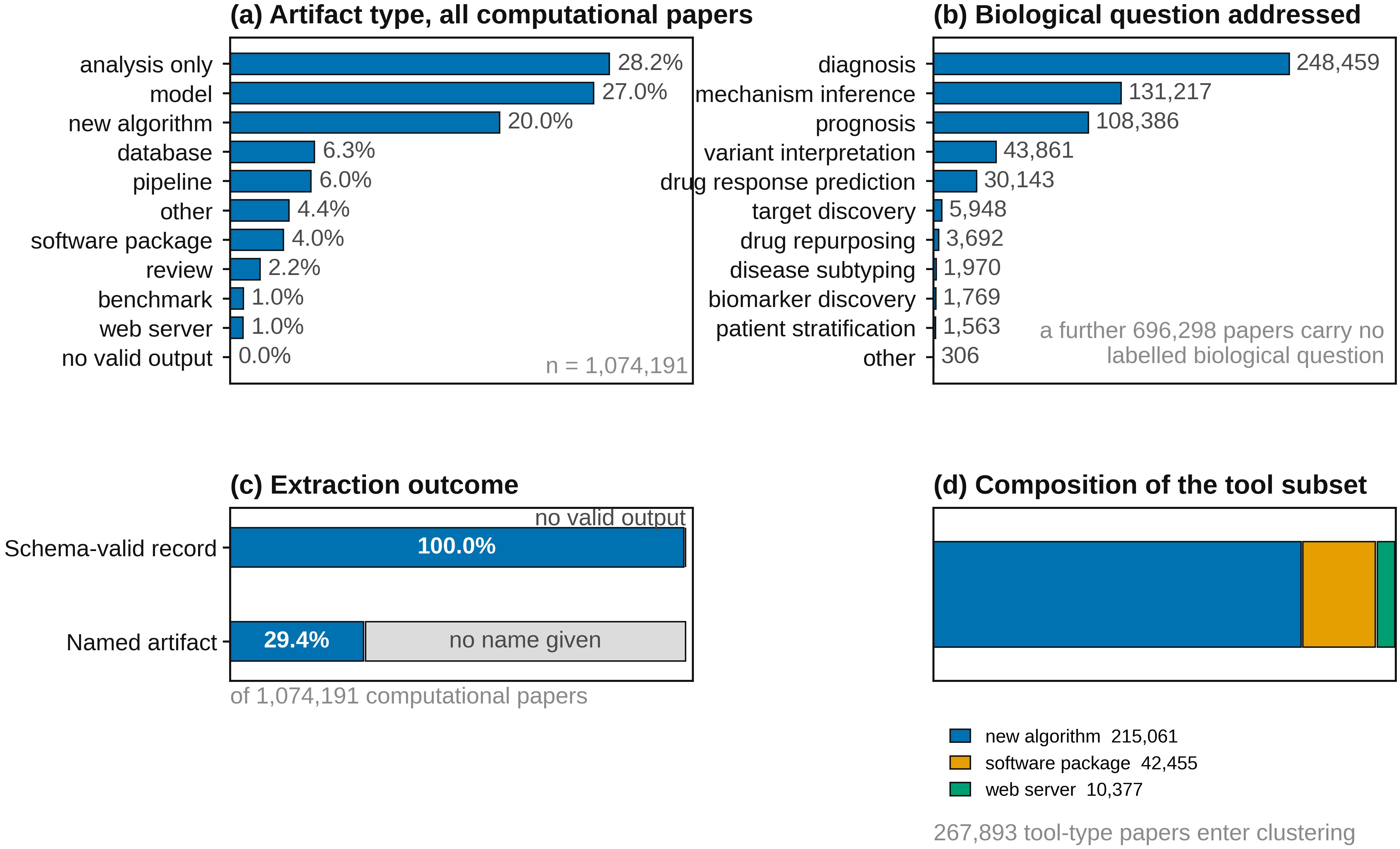}}

\textbf{Figure S1.} What the literature contributes. (a) Artifact types across all 1,074,191
papers screened as computational. (b) Biological questions addressed, as multi-label counts. (c) Extraction
outcome. (d) Composition of the tool-type subset by artifact type.

\end{document}